\documentclass[aip,jcp,reprint,amsmath,amssymb,floatfix,citeautoscript]{revtex4-1}
\usepackage{amsmath}
\usepackage[usenames,dvipsnames]{color}
\usepackage{amssymb}
\usepackage{braket}
\usepackage{graphicx}
\usepackage{ragged2e}
\usepackage[labelsep=period,format=plain,justification=justified,font=footnotesize]{caption}
\DeclareCaptionJustification{myjust}{\justifying}
\usepackage{times}
\usepackage{txfonts}

\begin{document}

\title{Approximate but Accurate Quantum Dynamics from the Mori Formalism: I. Nonequilibrium Dynamics}
\author{Andr\'{e}s Montoya-Castillo\footnote{Corresponding Author}}
\email{am3720@columbia.edu}
\affiliation{Department of Chemistry, Columbia University, New York, New York, 10027, USA}
\author{David R. Reichman}
\affiliation{Department of Chemistry, Columbia University, New York, New York, 10027, USA}

\date{\today}

\begin{abstract}
We present a formalism that explicitly unifies the commonly used Nakajima-Zwanzig approach for reduced density matrix dynamics with the more versatile Mori theory in the context of nonequilibrium dynamics.  Employing a Dyson-type expansion to circumvent the difficulty of projected dynamics, we obtain a self-consistent equation for the memory kernel which requires only knowledge of normally evolved auxiliary kernels.  To illustrate the properties of the current approach, we focus on the spin-boson model and limit our attention to the use of a simple and inexpensive quasi-classical dynamics, given by the Ehrenfest method, for the calculation of the auxiliary kernels.  For the first time, we provide a detailed analysis of the dependence of the properties of the memory kernels obtained via different projection operators, namely the thermal (Redfield-type) and population based (NIBA-type) projection operators.  We further elucidate the conditions that lead to short-lived memory kernels and the regions of parameter space to which this program is best suited. Via a thorough analysis of the different closures available for the auxiliary kernels and the convergence properties of the self-consistently extracted memory kernel, we identify the mechanisms whereby the current approach leads to a significant improvement over the direct usage of standard semi- and quasi-classical dynamics. 
\end{abstract}

\maketitle



\normalsize

\section{Introduction}
\label{Sec:Intro}

The continued effort to develop accurate and efficient approaches for the calculation of the dynamics of many-body quantum systems has produced a rich variety of methods, ranging from the numerically exact to the approximate.  While exact methods provide important benchmark results for model systems,\cite{Makri1992, Makri1995c, Makri1995d, Meyer1990,Beck2000,Wang2001a,Thoss2001,Ishizaki2009} their computational cost makes them impractical for realistic multidimensional systems.  Conversely, approximate methods, whether perturbative\cite{Bloch1957, Redfield1965, Leggett1987} or based on quasi-\cite{Gerber1982, Stock1995, Tully1971,Tully1990,Tully1998a} or semi-classical\cite{Meyer1979,Stock1997, Sun1998, Shi2003d, Miller1998, Miller2001a, Miller2009} approaches, tend to scale more gracefully with system size and can address both model and realistic systems, albeit at the expense of general applicability and accuracy.  For cases where one is interested in the dynamics of a limited number of degrees of freedom, the Nakajima-Zwanzig (NZ) equation\cite{Nakajima1958, Zwanzig1960a} provides a useful starting point for a plethora of methods based on generalized quantum master equations (GQME).  

 The NZ equation, which may be derived via the projection operator technique,\cite{Grabert1982, FickSauermann} dictates the evolution of the reduced density matrix (RDM) for the portion of the Hilbert space denoted as the system.  The influence of the complementary subspace, referred to as the bath, on the RDM dynamics appears in the form of a memory term, full knowledge of which is tantamount to solving the original problem.  The apparent simplicity of the NZ equation, however, belies the complexity of the memory term, which can be formidably difficult if not impossible to calculate exactly.  Despite the seeming conservation of difficulty, different treatments of the memory kernel have led to manageable and often very successful approximate and numerically  exact schemes.\cite{Redfield1965, Bloch1957, Leggett1987, Jang2011a, Tanimura1989, Tanimura2015}

A major difficulty in the calculation of the memory kernel lies in the fact that its dynamical evolution involves the ``projected'' propagator, $e^{i(1 - \mathcal{P})\mathcal{L}t}$, where $\mathcal{P}$ is the projector that defines the reduced dynamics. 
To circumvent the problem of projected dynamics, Shi and Geva\cite{Shi2003,Shi2004a,Zhang2006a} proposed a self-consistent expansion of the memory kernel, which requires the calculation of auxiliary kernels evolved with the \textit{normal} rather than the \textit{projected} propagator.  From an exact perspective, this approach is useful only if the numerical effort necessary for the calculation of the auxiliary memory kernels is less than that required for the direct calculation of the system dynamics. Using the numerically exact quasi-adiabatic path integral (QUAPI) method,\cite{Makri1995c,Makri1995d} Shi and Geva have shown that the memory kernel for the spin-boson model can decay up to 10 times faster than the system dynamics,\cite{Shi2003} lending credence to the feasibility of the self-consistent approach.  More recently, a similar scheme has been used by Rabani and co-workers within a path integral framework for the study of quantum transport problems.\cite{Cohen2011,Cohen2013,Wilner2013,Wilner2014a,Wilner2014} Just as importantly, applications of the method have successfully used semi- and quasi-classical theories to calculate the auxiliary kernels.\cite{Shi2004a}  In particular, Kelly, Markland, and coworkers have illustrated the impressive accuracy and robustness of this approach in both model and realistic problems.\cite{Kelly2013,Kelly2015, Pfalzgraff2015} These studies have led to two important conclusions: (i) the memory kernels are short-lived in a wide region of parameter space for canonical problems such as the spin-boson model and (ii) the self-consistent solution of the memory kernel using approximate dynamics can yield impressively more accurate results than \textit{direct} simulation of the RDM dynamics using the very same approximate method.


 Despite these important results, questions of general applicability still remain.  For instance, the conditions that lead to short-lived memory kernels remain unknown.  Further, it is still unclear how the breakdown of approximate methods (in unfavorable parameter regimes) affects the quality of the self-consistently extracted memory function. Perhaps most importantly, an understanding of the necessary and sufficient conditions from which the improvement in the accuracy of approximate dynamics from use in conjunction with the memory function formalism as observed in Refs. \onlinecite{Shi2004a,Kelly2013,Kelly2015} is lacking. Finally, the convergence properties of different versions of the auxiliary memory kernels arising from the alternative closures in the self-consistent expansion of the memory function have not yet been explored. We expect these convergence properties to differ when approximate methods are employed in the calculation of the auxiliary kernels. 

The remarkable utility of the NZ equation notwithstanding, objects beyond single-time nonequilibrium dynamics are either cumbersome to obtain within this framework.  Despite this difficulty, recent work has generalized the NZ equation to multi-time correlation functions.\cite{Ivanov2015}  In contrast, the more flexible Mori formulation permits direct extension to multiple-time and equilibrium correlation functions, which are essential in the treatment of linear\cite{NitzanBook} and non-linear spectroscopy,\cite{PrinciplesNonlinearSpec} and the calculation of chemical rate constants\cite{Yamamoto1960,Miller1998} and kinetic coefficients,\cite{MahanBook} to name a few examples.  For this reason, in paper I (this paper) of this series, we provide a unified Mori-type framework to approach single-time nonequilibrium correlation functions, and address several of the open questions listed above.  In a second paper, we present the a similar framework to treat equilibrium correlation functions.  It should be noted that a major advantage of the Mori formulation is that it can naturally address problems where no system-bath distinction exists, such as spin and fermion lattice models,\cite{Rasetti1991, Sachdev2011, Giamarchi2004} and quantum fluids.\cite{Feenberg1969, PinesNozieres1999, Poulsen2005, Markland2011}

The structure of this paper is as follows:  In Sec.~\ref{Sec:MoriApproach}, we present the formalism for nonequilibrium correlation functions from the Mori perspective and show that, with the appropriate choice of projection operator, one recovers equations identical to those arising from the conventional NZ treatment.  Sec. \ref{Sec:MoriApproach} introduces the spin-boson model and the proposes two types of projection operators for this model. Sec.~\ref{Sec:Closures} discusses different closures of the memory kernel based on different placements of the $\mathcal{Q} = 1 - \mathcal{P}$ projection operator and on the use of time derivatives.  To illustrate the arguments related to convergence, we implement the mean field Ehrenfest method, as first proposed in Ref.~\onlinecite{Kelly2015}, to obtain the auxiliary memory kernels.  We henceforth refer to the use of the Ehrenfest method coupled to the self-consistent extraction of the memory kernels as the GQME+MFT approach. In the Sec.~\ref{Sec:Results}, show the different properties of the memory kernels associated with the Redfield- and NIBA-type projectors, explore the performance of the GQME+MFT approach to SB models whose system-bath coupling is characterized by Ohmic and Debye spectral densities, and investigate the convergence properties of the distinct closures introduced in Sec.~\ref{Sec:Closures}.  In Sec.~\ref{Sec:Conclusions}, we conclude.

\section{Mori Approach}
\label{Sec:MoriApproach}

For illustrative purposes, we focus on the spin-boson (SB) Hamiltonian, which is representative of typical open quantum systems, but note that the current approach is general and may be applied to any Hamiltonian.\cite{Weiss} In particular, a major advantage of the Mori formalism developed here over the traditional NZ approach is the ability to treat systems with no natural system-bath separation, such as spin-chains and lattice models\cite{Rasetti1991, Sachdev2011, Giamarchi2004} or liquids.\cite{Feenberg1969, PinesNozieres1999, Poulsen2005, Markland2011}  We reserve these applications for later work.  

The SB Hamiltonian takes the form $H = H_S + H_B + V$. It contains a system part consisting of two sites offset by an energy bias 2$\varepsilon$ and with off-diagonal coupling $\Delta$, which is assumed to be independent of the bath coordinates, 
	\begin{equation}
	H_S = \varepsilon\sigma_z + \Delta\sigma_x,
	\end{equation}	 
	where $\sigma_i$ corresponds to the $i^{th}$ Pauli matrix. 
	The bath part of the Hamiltonian consists of independent harmonic oscillators, 	\begin{equation}
	H_B = \frac{1}{2} \sum_k \Big[\hat{P}_k^2 + \omega_k^2 \hat{Q}_k^2\Big],
	\end{equation}
	where $P_k$ and $Q_k$ are the mass weighted momenta and coordinates for the $k^{th}$ harmonic oscillator, respectively, and $\omega_k$ is the frequency of the $k^{th}$ mode. 
	The coupling between the system and bath is assumed to be linear in the bath coordinates and diagonal and antisymmetric with respect to the system basis, 
	\begin{equation}
	V = \alpha \sigma_z \sum_k c_k \hat{Q}_k,
	\end{equation}
	where $c_k$ is the coupling constant between the system and the $k^{th}$ oscillator, and $\alpha = \pm 1$ depending on the definition of the model.  The spectral density, $J(\omega)$, fully characterizes the system-bath interaction, and takes the form,
	\begin{equation}
	\begin{split}
	J(\omega) &= \alpha^2 \sum_k \frac{c_k^2}{2}\delta(\omega - \omega_k).
	\end{split}
	\end{equation}
	
	It is common to assume one of several forms for the spectral density. Two important cases describe Ohmic dissipation in condensed phase systems where $J(\omega)$ is proportional to $\omega$ as $\omega \rightarrow 0$.  These are the standard Ohmic spectral density\cite{Leggett1987} characterized by an exponential cutoff, and the Debye spectral density characterized by a Lorentzian cutoff:
	\begin{align} 
	J_{Oh}(\omega) &= \xi \omega e^{-\omega / \omega_c}, \label{Eq:OhmicSD}\\
	J_{De}(\omega) &= 2\lambda \omega_c \ \frac{\omega}{\omega^2 + \omega_c^2},\label{Eq:DebyeSD}
	\end{align}
where the  cutoff frequency $\omega_c$ determines the correlation time of the bath at sufficiently high temperatures.\cite{Aslangul1985} The reorganization energy, $\lambda = \pi^{-1} \int_0^{\infty} d\omega\ J(\omega)/\omega$, is a measure of the strength of the system-bath coupling.  In the case of the Ohmic spectral density, the Kondo parameter, $\xi = \pi \lambda / \omega_c$, is often used instead to gauge the coupling strength.

	To assess the applicability of the formalism presented here, we compare our results to exact nonequilibrium population dynamics of the SB model, 
	\begin{equation}\label{Eq:PopDynamics}
	\langle \sigma_z(t) \rangle = \mathrm{Tr}[\sigma_z(t) \ket{1}\bra{1} \rho_B], 
	\end{equation}
	where $\rho_B = e^{-\beta H_B} / \mathrm{Tr}_B[e^{-\beta H_B}]$ is the equilibrium density operator for the uncoupled bath, $\beta = [k_BT]^{-1}$ is the inverse thermal energy, and the initial condition $\ket{1}\bra{1}\rho_B$ corresponds to a Frank-Condon transition.

\subsection{Generalized Nakajima-Zwanzig-Mori Equation}
\label{Subsec:GeneralizedNZMEq}
Here, we deviate from the derivations commonly given for the NZ equation or the Mori equation of motion (EOM) for an operator and instead focus on a generalized EOM for the full propagator, 
	\begin{equation} \label{Eq:GeneralNZMEq}
	\begin{split}
	\frac{d}{dt}e^{i\mathcal{L}t} &= ie^{i\mathcal{L}t}\mathcal{P}\mathcal{L} + i \mathcal{Q}e^{i\mathcal{L}\mathcal{Q}t}\mathcal{L} \\
	&\qquad -  \int_0^{t}d\tau\ e^{i\mathcal{L}(t-\tau)}\mathcal{P}\mathcal{L}\mathcal{Q}e^{i\mathcal{L}\mathcal{Q}\tau}\mathcal{L}
	\end{split}
	\end{equation}
where $\mathcal{P}$ is the projection operator that defines the subsystem whose dynamics we seek and $\mathcal{Q} = 1 - \mathcal{P}$ is the complementary projection operator. This equation is general and can be employed within both the NZ and Mori approaches.  For instance, applying Eq.~(\ref{Eq:GeneralNZMEq}) on an operator $\hat{A}$, yields the Mori EOM for that operator. Conversely, taking the Hermitian conjugate of Eq.~(\ref{Eq:GeneralNZMEq}), applying it on the initial density of the system and bath, $\rho_0$, and acting the projection operator $\mathcal{P}$ from the left followed by a trace over the bath degrees of freedom yields the NZ equation for the system's RDM.  


Essentially all work on the self-consistent expansion of the memory kernel has employed the thermal (Argyres-Kelley) projection operator $\mathcal{P} = \mathcal{R}_{B} \mathrm{Tr}_B[...]$,\cite{Shi2003, Shi2004a, Zhang2006a, Cohen2011, Cohen2013, Kelly2013, Kelly2015} where $\mathcal{R}_B$ is a bath operator with unit trace and $\mathrm{Tr}_{B}[...]$ corresponds to partial trace over the bath.  To use the Mori formulation, it is convenient to rewrite the thermal projector, in the Heisenberg picture using Liouville notation,\footnote{For an introduction to this notation, see Chapter 2 in Ref. \onlinecite{PrinciplesNonlinearSpec}} as $\mathcal{P} = \sum_i\ket{A_i}\rangle \langle\bra{\mathcal{R}_{Bi} A_i}$, where $\{A_i\}$ contains all outer product states spanning the system.  For the spin-boson model, $A_i \in \{\ket{1}\bra{1}, \ket{2}\bra{1}, \ket{1}\bra{2}, \ket{2}\bra{2} \}$.  

Using the thermal type projector, applying Eq.~(\ref{Eq:GeneralNZMEq}) to $\ket{A_k}\rangle$, and closing on the left with $\langle \bra{\mathcal{R}'_{Bj} A_j}$, where $\mathcal{R}'_{Bj}$ is again a bath operator with unit trace that may be different from $\mathcal{R}_{Bj}$ in the projection operator, yields the following set of EOMs for system observables, 
	\begin{equation} \label{Eq:MoriSBmodel}
	\begin{split}
	\frac{d}{dt}\mathcal{C}(t) &= i\mathcal{C}(t)\mathcal{X} + \mathcal{I}(t) -  \int_0^{t}d\tau\ \mathcal{C}(t-\tau)\mathcal{K}(\tau), 
	\end{split}
	\end{equation}
where $\mathcal{X}_{jk} = \langle\bra{\mathcal{R}_{Bj} A_j} \mathcal{L} \ket{A_k}\rangle$ is a static rotation matrix, $\mathcal{C}_{jk}(t) = \langle \bra{\mathcal{R}'_{Bj} A_j} e^{i\mathcal{L}t} \ket{A_k}\rangle = \mathrm{Tr}[(\mathcal{R}'_{Bj})^{\dagger} A_j^{\dagger} A_k(t)]$ corresponds to nonequilibrium averages of populations and coherences with all possible factorizable initial conditions, and $\mathcal{I}_{jk}(t) =i \langle\bra{\mathcal{R}'_{Bj} A_j}  \mathcal{Q}e^{i\mathcal{L}\mathcal{Q}t}\mathcal{L}\ket{A_k}\rangle$ is the so-called the inhomogeneous term.  The elements of the memory kernel are given by,
	\begin{align}
	\mathcal{K}_{jk}(t) &= \langle \bra{\mathcal{R}_{Bj} A_j} \mathcal{L}\mathcal{Q}e^{i\mathcal{Q}\mathcal{L}t}\mathcal{Q}\mathcal{L} \ket{A_k}\rangle. \label{Eq:MemoryKernel2}
	\end{align}
When $\mathcal{R}_{Bj}' = \mathcal{R}_{Bj}$, the inhomogeneous term disappears, $\mathcal{I}(t) = 0$.  Often, $\mathcal{R}_{Bj} = R_{Bj}/\mathrm{Tr}_B[R_{Bj}]$ is chosen such that $ R_{Bj} \in \{ e^{-\beta H_B} ,  e^{-\beta (H_B \pm \sum_k c_k \hat{Q}_k)} \}$, which correspond to the harmonic oscillator bath at equilibrium with the ground electronic state or with one of the two excited states, respectively.  Initial conditions of the form $\rho(0) = \rho_S(0) \mathcal{R}_{Bi}$, where $\rho_S(0)$ is an arbitrary system operator and $\mathcal{R}_{Bi}$ is taken from the set above, correspond either to a Frank-Condon excitation where the bath is in the electronic ground state also called the spectroscopic initial condition, or a charge transfer initial condition where the bath is in equilibrium with one of the excited states. 

We henceforth refer to the thermal projector above with the additional restriction that $R_{Bi} = \exp[-\beta H_B]$ as Redfield-type.\cite{Bloch1957,Redfield1965}  The reason for this name is that truncation of the memory kernel at second order in $\mathcal{Q}$ is equivalent to a second-order perturbative expansion of the memory kernel with respect to the system-bath coupling, which corresponds to Redfield theory.  We further note that the choice for $R'_{B,i}$ remains flexible.  

  An important alternative for the projection operator consists of restricting the set $\{ A_i\}$ to the system populations $B_i = \ket{i}\bra{i}$, and choosing $R_{Bi} = \exp[- \beta(H_B + (-1)^i  \sum_k c_k \hat{Q}_k)]$.  When using this projection operator, a similar second-order truncation of the memory kernel with respect to $\mathcal{Q}$ leads to equations that are equivalent to the non-interacting blip approximation (NIBA), which is a second order expansion in the electronic coupling $\Delta$ as opposed to the system-bath coupling.\cite{Leggett1987, Sparpaglione1988a}  Accordingly, we hereafter refer to this projector as NIBA-type.  As in the case of the Redfield-type projector, the choice for $R'_{B,i}$ determines whether the inhomogeneous term is zero or finite.

\section{Self-Consistent Expansions for $\mathcal{K}(t)$}
\label{Sec:Closures}

As mentioned in Sec.~\ref{Sec:Intro}, the main difficulty associated with the memory kernel, Eq.~(\ref{Eq:MemoryKernel2}), is the presence of the projected propagator, $e^{i\mathcal{L}\mathcal{Q}t}$.  To circumvent this problem, Shi and Geva proposed the use of the Dyson identity, 
	\begin{align}
	e^{(A+B)t} &= e^{At} + \int_0^t ds\  e^{As}Be^{(A+B)(t - s)} \label{Eq:DysonEq1}\\
	&= e^{At} + \int_0^t ds\  e^{(A+B)s}Be^{A(t - s)},\label{Eq:DysonEq2}
	\end{align}
yielding a self-consistent expansion of the memory kernel that only involves unprojected dynamics.\cite{Shi2003,Zhang2006a}  Not surprisingly, Eq.~(\ref{Eq:GeneralNZMEq}) can also be derived using the Dyson identity, Eqs.~(\ref{Eq:DysonEq1}) and (\ref{Eq:DysonEq2}).   Despite considering only time-independent Hamiltonians in this work, extension of the formalism to time-dependent Hamiltonians is simple and only requires the time-ordered form for the propagators in Eqs.~(\ref{Eq:DysonEq1}) and (\ref{Eq:DysonEq2}).  

It is important to remark that in the literature, different variants of the Dyson expansion have led to a menagerie of seemingly distinct expressions, which differ with respect to the number of and type of auxiliary kernels employed.\cite{Zhang2006a}   When these distinct expressions are evaluated via exact methods, all expansions yield equivalent results, up to numerical errors. However, when the auxiliary kernels are computed via approximate methods, different expansions can lead to memory kernels with different properties.  In the following, we show that there is only a limited number of integral equation forms that can yield numerically distinct approximate memory kernels from the self-consistent solution of the resulting integral equations.  

\subsection{Bare expansions: Backward and Forward $\mathcal{Q}$}
\label{Subsec:BareExp}

The expansion of the memory kernel, Eq.~(\ref{Eq:MemoryKernel2}), takes the following form, 
	\begin{equation}\label{Eq:SCKb}
	\begin{split}
\mathcal{K}(t) &= \mathcal{K}^{(1)}(t) + \int_0^t d\tau\ \mathcal{K}^{(3b)}(t-\tau)\mathcal{K}(\tau),
	\end{split}
	\end{equation}
	where the superscript $b$ refers to the placement of the $\mathcal{Q}$ in the projected propagator as ``backward'' with respect to the placement of the Liouvillian, i.e., $e^{i\mathcal{Q}\mathcal{L}t}$. The auxiliary memory kernels take the forms
	\begin{align}
	[\mathcal{K}^{(1)}(t)]_{jk} &= \langle \bra{\mathcal{R}_{Bj} A_j} \mathcal{L}\mathcal{Q}e^{i\mathcal{L}t}\mathcal{Q}\mathcal{L} \ket{A_k}\rangle , \label{Eq:K1b1}\\
	[\mathcal{K}^{(3b)}(t)]_{jk} &= -i\ \langle \bra{\mathcal{R}_{Bj} A_j} \mathcal{L}\mathcal{Q}e^{i\mathcal{L}t}  \ket{A_k}\rangle \label{Eq:K3b1}. 
	\end{align}

A seemingly distinct type of closure that is commonly used in the literature involves a \textit{third} auxiliary memory kernel. In Appendix \ref{App:Closures}, we show that the three-member expansions are equivalent to the two-member expansions given by Eqs.~(\ref{Eq:SCKb}) and (\ref{Eq:SCKf}).  We further note that the auxiliary kernels we obtain are equivalent to those used by others in the field.\cite{Shi2003,Kelly2013,Kelly2015} 

A second set of closures makes use of the identity $e^{i\mathcal{Q}\mathcal{L}t}\mathcal{Q} = \mathcal{Q}e^{i\mathcal{L}\mathcal{Q}t}$.  To indicate that the $\mathcal{Q}$ is to the right of the Liouvillian $\mathcal{L}$ in the propagator, we have used the superscript $f$ (indicating a ``forward'' placement).  Inserting the previous identity in Eq.~(\ref{Eq:MemoryKernel2}) leads to the following expansion upon insertion of the Dyson identity, 
	\begin{equation}\label{Eq:SCKf}
	\begin{split}
\mathcal{K}(t) &= \mathcal{K}^{(1)}(t) + \int_0^t d\tau\ \mathcal{K}(t-\tau)\mathcal{K}^{(3f)}(\tau).  
	\end{split}
	\end{equation}
We note that $\mathcal{K}^{(1)}(t)$ has the form given in Eq.~(\ref{Eq:K1b1}), and $\mathcal{K}^{(3f)}(t)$ has the following form,
	\begin{align}
	[\mathcal{K}^{(3f)}(t)]_{jk} &= -i\ \langle \bra{\mathcal{R}_{Bj} A_j} e^{i\mathcal{L}t} \mathcal{Q} \mathcal{L}\ket{A_k}\rangle \label{Eq:K3f1}. 
	\end{align}
It bears remarking that Eqs.~(\ref{Eq:SCKf}) and (\ref{Eq:SCKb}) differ in the placement of $\mathcal{K}(t)$ under the integral and Eqs.~(\ref{Eq:K3f1})  and (\ref{Eq:K3b1}) differ in whether $\mathcal{Q}\mathcal{L}$ or its Hermitian conjugate act on operators that require sampling only at $t = 0$ or at finite times. 

\subsection{Expansions using time-derivatives}
\label{Subsec:TimeDerivExp}

Because the auxiliary kernels given by Eqs.~(\ref{Eq:K1b1}) and (\ref{Eq:K3b1}) or (\ref{Eq:K3f1}) require sampling of additional bath operators at $t=0$ and at finite times, convergence of these functions, at least within the context of semi- and quasi-classical methods, necessitates the sampling of a larger number of bath realizations than for the direct simulation of $\mathcal{C}(t)$, making the initial step of the calculation more expensive, even if trivially parallelizable. To avoid this added complexity and expense, the expressions for the auxiliary kernels can be rewritten as time derivatives of simpler correlation functions, including $\mathcal{C}(t)$ itself. Indeed, this observation has been made in recent work.\cite{Cohen2011,Cohen2013,Wilner2013, Wilner2014, Wilner2014a, Kidon2015a}

Here we focus on three types of auxiliary memory kernels that exploit different placements of the time-derivative.  The first type replaces the Liouvillian acting on operators that require dynamic sampling and leaves the Liouvillian acting on the static parts intact.  In this scheme, $\mathcal{K}^{(3b)}(t)$ remains unchanged.  The other auxiliary memory kernels may be expressed as follows,
	\begin{align}
	\mathcal{K}^{(1)}_{1}(t) &= \dot{\mathcal{K}}^{(3b)}(t) - i \mathcal{K}^{(3b)}(t) \mathcal{X}, \label{Eq:K1b1ActDeriv}\\
	\mathcal{K}^{(3b)}_{1}(t) &=  \mathcal{K}^{(3b)}(t)\label{Eq:K3b1ActDeriv},\\
	\mathcal{K}^{(3f)}_{1}(t) &= -\dot{\mathcal{C}}(t) + i \mathcal{C}(t)\mathcal{X}. \label{Eq:K3f1ActDeriv}
	\end{align}

The second type focuses on replacing the Liouvillian operating on static operators with the time derivative, yielding the following expressions, 
	\begin{align}
	\mathcal{K}^{(1)}_{2}(t) &= \dot{\mathcal{K}}^{(3f)}(t) - i\mathcal{X} \mathcal{K}^{(3f)}(t) , \label{Eq:K1b1PasDeriv}\\
	\mathcal{K}^{(3b)}_{2}(t) &= -\dot{\mathcal{C}}(t)+ i \mathcal{X}\mathcal{C}(t)\label{Eq:K3b1PasDeriv},\\
	\mathcal{K}^{(3f)}_{2}(t) &=  \mathcal{K}^{(3f)}(t). \label{Eq:K3f1PasDeriv}
	\end{align}
	
The final type replaces all Liouvillians with time derivatives, 
	\begin{align}
	\mathcal{K}^{(1)}_{3}(t) &= -\ddot{\mathcal{C}}(t) + i\{ \dot{\mathcal{C}}(t), \mathcal{X} \} -  \mathcal{X}\mathcal{C}(t)\mathcal{X}, \label{Eq:K1b1DDeriv}\\
	\mathcal{K}^{(3b)}_{3}(t) &=  \mathcal{K}^{(3b)}_2(t)\label{Eq:K3b1DDeriv},\\
	\mathcal{K}^{(3f)}_{3}(t) &= \mathcal{K}^{(3f)}_1(t). \label{Eq:K3f1DDeriv}
	\end{align}

It should be noted that Eqs.~(\ref{Eq:K1b1ActDeriv})--(\ref{Eq:K3f1DDeriv}) are exact identities in the context of exact quantum dynamics but \textit{may yield different results when approximate quantum dynamics are employed.}  

For clarity in the subsequent discussion, we henceforth refer to the different closures via abbreviations of the form $c(xy)$, where $x \in \{ \mathrm{f(orward)},\   \mathrm{ b(ackward)} \}$, and $y \in \{0,1,2,3\}$ where $0$ denotes the bare expansion and 1, 2, 3 the three types of expansion in the present section.  For example, the $cb2$ closure refers to the use of Eqs.~(\ref{Eq:K1b1PasDeriv}) and (\ref{Eq:K3b1PasDeriv}) to solve for $\mathcal{K}(t)$ in Eq.~\ref{Eq:SCKb}.

\section{Results}
\label{Sec:Results}

  The recent success achieved in using semi- and quasi-classical schemes to calculate the auxiliary kernels required in the self-consistent extraction of memory kernels underscores the importance of understanding the properties of this program in more detail. Consequently, we employ a simple quasi-classical method, namely Ehrenfest dynamics,\cite{Tully1998a,Grunwald2009} to obtain the auxiliary memory kernels and study the performance of the Redfield- and NIBA-type projectors and of the different closures available for the kernels.  The procedural steps we follow can be summarized as follows: 
\begin{enumerate}
\item Calculate the various auxiliary kernels via a dynamical method of choice.  Here, we use the approximate Ehrefest approach. 

\item Solve Eqs.~(\ref{Eq:SCKb}) or (\ref{Eq:SCKf}) iteratively until the relative error becomes negligible.  Our threshold is $10^{-10}$. 

\item Numerically integrate Eq.~(\ref{Eq:MoriSBmodel}) subject to the appropriate initial conditions.  In the following we use a second-order Runge-Kutta algorithm.  
\end{enumerate}

The memory kernel decays to zero for a large sections of parameter space for the SB and other impurity-type models.\footnote{We are currently exploring cases for which the memory kernel decays to zero very slowly or decays to a finite constant.}  We refer to the timescale that determines this decay as the memory lifetime. As mentioned in the Introduction, the computational efficiency of the memory function approach depends sensitively on this lifetime.   

When approximate methods, such as semi- and quasi-classical schemes, are used to calculate the auxiliary kernels, the extracted memory function can accrue errors that are expected to grow with increasing simulation time.  Hence, the decay of the memory kernel may not be accurately captured by these methods.  Previous applications of the memory function approach have implemented a cutoff time for the memory kernel, after which all its components are set to zero.  Sensitivity of the results to this cutoff time will be discussed in Sec.~\ref{Subsec:ConvergenceAndDynamics}.  For all results showing only one GQME+MFT curve, the cutoff time, $\tau_c$, was chosen at the point where the extracted $\mathcal{C}(t)$ dynamics reached a plateau of stability. 
For a more thorough discussion of the Ehrenfest method and its implementation for obtaining the auxiliary kernels, see Appendices \ref{App:ICEhrenfest} and \ref{App:EhrenfestCFs}.   

\subsection{Projection Operators}
\label{Subsec:ProjectionOperators}

In this section, we restrict our attention to the Ohmic spectral density, a model whose performance has already been studied extensively using the Redfield-type projection operator and closure scheme $cb0$ by Kelly, Markland, and co-workers.\cite{Kelly2015}  The purpose of this section is mainly to provide an analysis of the NIBA-type memory kernels and show the viability of the GQME+MFT approach using both the Redfield- and NIBA-type projectors.  Here and in Sec. \ref{Subsec:Debye}, we show results only for the $cb1$ closure, and postpone the discussion of the closure dependence of the results to Sec.~\ref{Subsec:ConvergenceAndDynamics}.   

\begin{figure}[t]
\includegraphics[width=8.5cm]{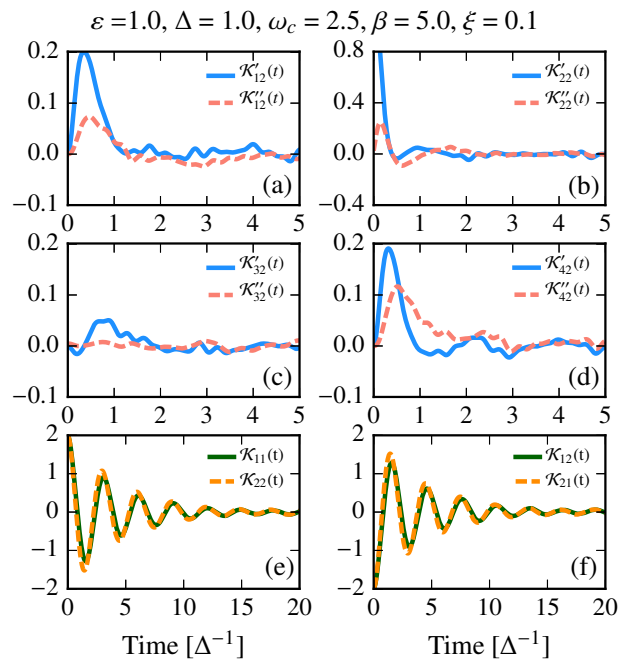} 
\caption{Redfield- and NIBA-type $cb1$ memory kernels for the SB model characterized by the Ohmic spectral density.  Panels (a)-(d) correspond to the real (solid) and imaginary (dashed) parts of the Redfield-type memory kernel elements $\mathcal{K}_{x2}(t) = \mathcal{K}'_{x2}(t) + i\mathcal{K}''_{x2}(t)$.  Panels (e)--(f) display \textit{all} components of the NIBA-type memory kernel, $\mathcal{K}(t)$.}\label{Fig:Red_KMFig3b_Kall}
\end{figure}

\begin{figure}[b]
\includegraphics[width=8.5cm]{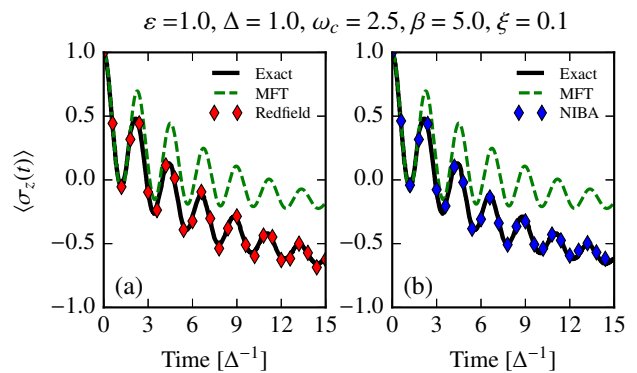} 
\caption{Population dynamics for the realization of the spin-boson model corresponding to the parameters in Fig.~\ref{Fig:Red_KMFig3b_Kall}.  Panel (a) compares the dynamics obtained from the Redfield-type memory kernel with a cutoff time $\tau_c = 2.0$ with the standard MFT and exact dynamics.  Panel (b) provides a similar comparison with the dynamics obtained from the NIBA-type memory kernel, with a cutoff time of $\tau_c = 15$. Only results for the $cb1$ closure are shown. Exact results for the Ohmic SB model ($\alpha = -1$) are obtained from Ref. \onlinecite{Kelly2015}.}\label{Fig:KMFig3b_BestPopulations}
\end{figure}


 We first compare the different properties of the Redfield- and NIBA-type memory kernels for a realization of the biased ($\varepsilon = 1$) spin-boson model characterized by weak system-bath coupling ($\xi = 0.1$), low temperature ($\beta = 5.0$), and a moderately fast bath ($\omega_c = 2.5$). Fig.~\ref{Fig:Red_KMFig3b_Kall} shows a representative set of components of the Redfield-type memory kernel, $\mathcal{K}_{x2}(t)$, in panels (a)-(d), and all components of the NIBA-type memory kernel $\mathcal{K}(t)$ in panels (e)-(f).  Comparison of the two types of type memory kernel reveals the different timescales associated with their decay.  Although seemingly noisy at longer times, the Redfield-type memory kernel has a short lifetime ($\tau_c \sim 2$), while the NIBA-type memory kernel decays much more slowly, $(\tau_c > 15)$.  

Fig.~\ref{Fig:KMFig3b_BestPopulations} illustrates the dynamics for the parameters used in Fig.~\ref{Fig:Red_KMFig3b_Kall}. Despite capturing the correct oscillation frequency and amplitude decay, the Ehrenfest dynamics (green dashes) fails to capture the long-time limit of the populations.  Indeed, because of the assumption of a classical bath, the Ehrenfest method is known to violate detailed balance.\cite{Parandekar2006} Instead, the dynamics resulting from both the Redfield- and NIBA-type GQMEs with the $cb1$ closure quantitatively agree (to within graphical accuracy) with the exact results, showing that either method presented here is viable for recovering highly accurate dynamics from approximate dynamics.  

\begin{figure}[t]
\includegraphics[width=8.5cm]{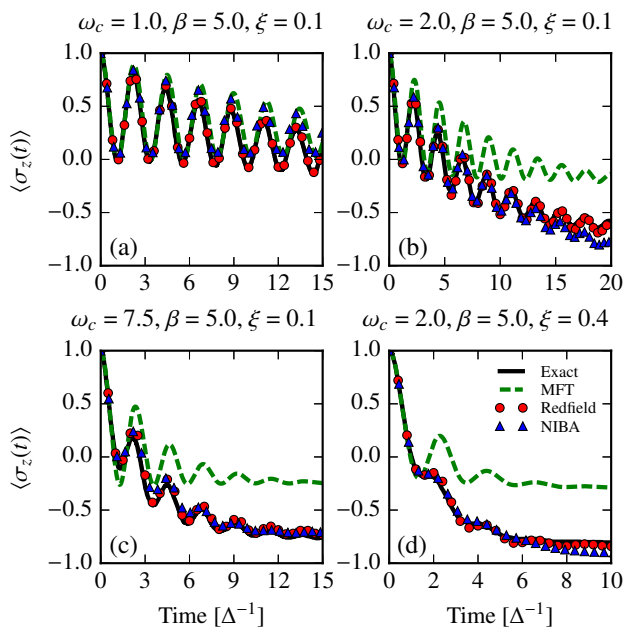} 
\caption{Population dynamics for four sets of parameters for the spin-boson model, assuming closure 1b.  For all panels $\varepsilon = 1$ and all parameters are in units of $\Delta$.  The results shown correspond to the $cb1$ closure. Exact results are obtained from Ref. \onlinecite{Kelly2015}.}\label{Fig:OhmicPopPanels}
\end{figure}

To understand the difference in the lifetimes of the Redfield- and NIBA-type memory kernels we recall that the Mori approach to Brownian motion,\cite{Mori1965} which focuses on the properties of a massive particle suspended in a bath of lighter particles, relies on the separation of timescales for the dynamics of the heavy and light particles.  This separation of timescales is made effective via the projection operator, which must be chosen such that it contains all slow variables associated with the massive particle.  Appropriate inclusion of all slow variables in the projector ensures that the memory kernel decays on a shorter timescale than the system dynamics.\cite{Reichman2005} The Redfield-type projector spans the entire Hilbert space of the system, whereas the NIBA-type projector excludes projections onto coherences, $\ket{i}\bra{j}$ where $i \neq j$.  While coherences often decay faster than populations, their decay is often slower than bath correlations, as long as the bath dimensionality is large. Hence, the slower time-scale associated with the decay of the coherences induces the slow decay of the NIBA-type memory kernels. This conclusion further suggests that the NIBA-type projector may be most useful in instances of fast system relaxation, such as strong system-bath coupling cases at high temperatures.  

It has been suggested that the success of the memory function program, where the auxiliary kernels are calculated via semi- and quasi-classical schemes like Ehrenfest mean-field theory,\cite{Kelly2015} the momentum-jump solution to the quantum-classical Liouville equation,\cite{Kelly2013} and the linearized semiclassical initial value representation (LSC-IVR) scheme,\cite{Shi2004} relies primarily on the confluence of two important factors.  First, the memory kernels are short-lived in comparison to the desired system dynamics.  Second, approaches based on semi-classical arguments are more accurate at short times.  Considered in tandem, these factors imply that the present scheme can lead to highly accurate short-time memory kernels, thus avoiding problems associated with the long-time dynamics produced by these approximate methods by virtue of fast memory decay. However, the ability of the slowly decaying NIBA-type memory kernel to nearly recover the exact dynamics raises an important question: given the long lifetime of the NIBA-type kernel, how can it remain sufficiently accurate at long times to correct the long-time behavior of the bare quasi-classical dynamics?  To answer this question, it is necessary to  scrutinize the form of the auxiliary kernels.  The NIBA-type auxiliary kernels for closure $cb1$ include two types of correlation functions, $q^{(00)}_{nm}(t)$ and $q^{(10a)}_{nm}(t)$ given by Eqs.~(\ref{Eq:q00}) and (\ref{Eq:q10p}). Clearly, $q^{(00)}_{nm}(t)$ is the Ehrenfest version of $\mathcal{C}(t)$, while $q^{(10a)}_{nm}(t)$ involves a new type of correlator that requires the sampling of an additional bath operator, $\zeta^W =  - \alpha \sum_j c_j P_j \tanh(\beta \omega_j / 2) / \omega_j$, at $t=0$. At this point, two possible reasons for the improvement afforded by the NIBA-type approach seem likely.  First, it may be that the Ehrenfest method describes the the dynamics of coherences, which are required as input in the auxiliary kernels, better than those of populations.  Second, $q^{(10a)}_{nm}(t)$ contains exact sampling of the bath operator $\zeta^W$ at $t = 0$, which may encapsulate important information about the system-bath interaction. With the information above, however, it is difficult to decide on which hypothesis is more likely.  We return to this discussion in Sec.~\ref{Subsec:ConvergenceAndDynamics}. The previous questions notwithstanding, the ability of the long-lived NIBA-type memory kernel to produce dynamics that are comparably accurate to those obtained via the Redfield-type approach underscores the fact that \textit{a rapidly decaying memory function is not required for the success of the GQME+MFT approach}. 
 

Fig.~\ref{Fig:OhmicPopPanels} provides a more thorough test of the performance of the NIBA-type projector.  Here we compare the NIBA-type GQME+MFT dynamics to exact results for the cases addressed by Kelly \textit{et al.}\cite{Kelly2015} in their recent work characterizing the performance of the Redfield-type GQME+MFT approach.  For convenience, we include the results from the Redfield-type projector as well. We further remark that the focus of these cases is the performance of the present approach to biased systems coupled weakly ($\lambda = \pi \xi / \omega_c < 1$) to a bath characterized by varying timescales, evident in the range of $\omega_c$.  As is clear from Fig.~\ref{Fig:OhmicPopPanels}, direct use of the Ehrenfest MFT method consistently leads to incorrect long-time values of the population difference.  In agreement with the work of Kelly and co-workers,\cite{Kelly2015} the Redfield-type GQME+MFT method quantitatively corrects the dynamics in all considered cases. The NIBA-type approach generally provides clear improvement over direct use of MFT, but is slightly less accurate than the Redfield-type GQME+MFT.  Regardless, the improvement of the dynamics produced by the NIBA-type projector is remarkable not just because of the fact that the memory function is long-lived, but also because such an approach is not tailored for the weak system-bath coupling limit as is the Redfield-type projector where the benchmark calculations of Fig.~\ref{Fig:OhmicPopPanels} have been performed. 

\subsection{Debye Spectral Density}
\label{Subsec:Debye}

Due to its slower decay at large frequencies, the Debye spectral density is generally considered a more challenging case for trajectory-based dynamical methods.\cite{Berkelbach2012} Here we show that the conclusions drawn from the Ohmic case, namely that the GQME+MFT method can significantly improve the problematic MFT dynamics for weakly coupled, biased systems at low temperatures, are similarly applicable to the Debye case.  A few differences are worth mentioning, chief among them that the highly oscillatory nature of the Redfield-type memory kernels for this spectral density generally means that a larger number of trajectories to achieve convergence are required. 

\begin{figure}[t]
\includegraphics[width=8.5cm]{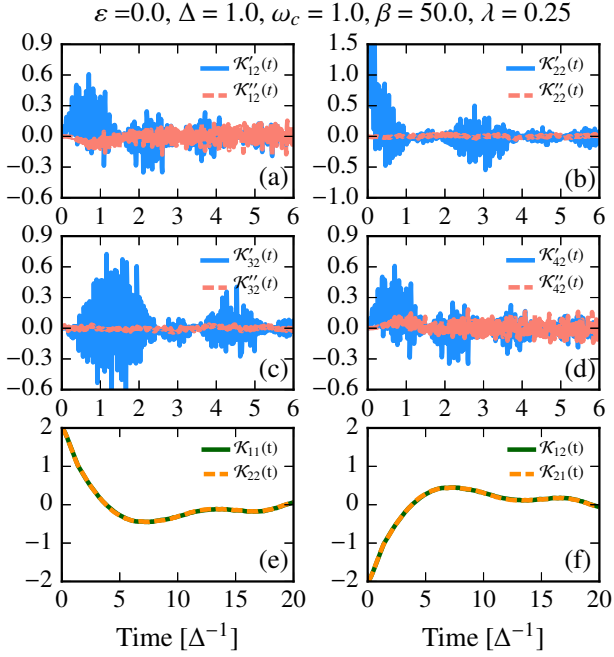}
\caption{Redfield- and NIBA-type memory kernels using the $cb1$ closure for the unbiased SB model characterized by the Debye spectral density.  Panels (a)--(d) correspond to the real (solid) and imaginary (dashed) parts of the Redfield-type memory kernel elements $\mathcal{K}_{x2}(t)$.  Panels (e)-(f) display \textit{all} components of the NIBA-type memory kernel. }\label{Fig:Debye_K}
\end{figure}

Panels (a)-(d) of Fig.~\ref{Fig:Debye_K} show the components $\mathcal{K}_{x2}(t)$ for the Redfield-type memory kernel, while panels (e)-(f) show all components of the NIBA-type memory kernel. While the NIBA-type memory kernels do not show any obvious differences from their Ohmic counterparts, the Redfield-type memory kernel displays recurrent beating alongside overall decay. The presence of this much stronger oscillatory behavior requires that the dynamics of high frequency modes in the Ehrenfest procedure be treated more accurately than is necessary in the Ohmic case. We discuss this issue in more depth in the next section, where we explore the convergence properties of the difference closures.  

\begin{figure}[t]
\includegraphics[width=8.5cm]{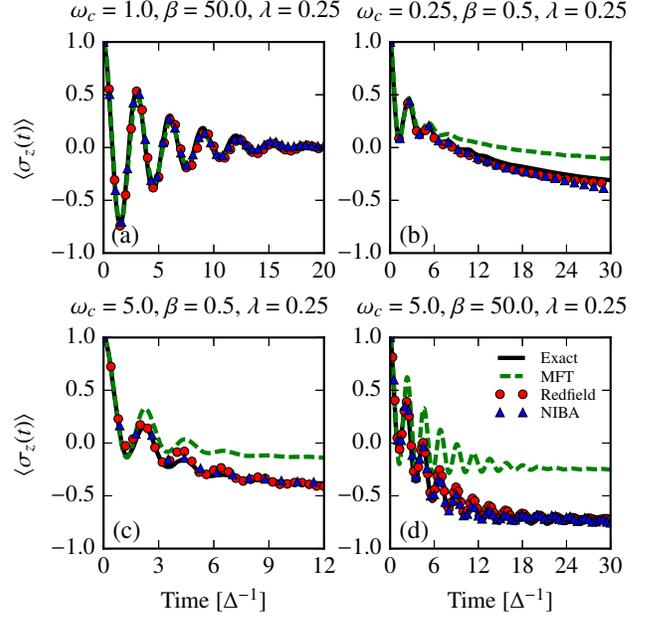} 
\caption{Population dynamics calculated from the $\mathcal{Q}$-forward closures of the memory kernel.  For this set of parameters the $\mathcal{Q}$-backward closures yield equivalent results. The results shown correspond to the $cb1$ closure. Exact dynamics for the Debye SB model ($\alpha = 1$) are obtained from Ref. \onlinecite{Thoss2001}.}\label{Fig:DebyePopPanels}
\end{figure}

Fig.~\ref{Fig:DebyePopPanels} presents some illustrative examples of the capability of the GQME+MFT approach to yield accurate dynamics for the biased SB model at low temperatures, over a wide range of $\omega_c$. Panel (a), which corresponds to an \textit{unbiased} case characterized by weak system-bath coupling at low temperature ($\beta = 50.0$) and an intermediate bath frequency ($\omega_c = 1.0$), shows nearly perfect agreement between the Ehrenfest and exact dynamics.  Both the Redfield- and NIBA-type GQME+MFT approaches are able to recover the remarkable agreement between the Ehrenfest and exact dynamics.  Panels (b)-(d) correspond to biased cases, spanning a wide range of bath frequencies ($\omega_c =$ 0.25, 5.0) and temperatures ($\beta = $0.5, 50.0).  As expected, the bare Ehrenfest method leads to incorrect long-time limits for all three biased cases.  As in the Ohmic case, both the Redfield- and NIBA-type approaches yield results in almost quantitative agreement with the exact dynamics.  Slight deviations are evident, as in panel (b), where the NIBA-type GQME slightly underestimates the long-time limit of the population difference. Perhaps the most difficult case for the current approach, panel (d), shows that the NIBA-type GQME+MFT treatment leads to overly damped oscillations at long times, whereas the Redfield-type approach yields results in near quantitative agreement with the exact dynamics.  In short, the results in Fig.~\ref{Fig:DebyePopPanels} illustrate the robustness of the approach for weak-coupling cases over a wide range of bath frequencies and temperatures.  


\subsection{Memory Kernel Closures and Dynamics}
\label{Subsec:ConvergenceAndDynamics}

In Sec. \ref{Sec:Closures}, we introduced \textit{eight} different closures of the memory kernels.  These include two subsets consisting of the $\mathcal{Q}$-forward and $\mathcal{Q}$-backward closures, which are further subdivided into the bare expansion ($cf0$ and $cb0$) and three expansions that use numerical derivatives of the simulated correlation functions, ($cf1$--$cf3$ and $cb1$--$cb3$). While the resulting dynamics do not differ when the auxiliary kernels are calculated via exact methods, the same claim is not necessarily true when using approximate dynamics. Here, we continue to use the Ehrenfest method to illustrate the sensitivity of the results that occur across the spectrum of closures.

 To inform the discussion on the properties of different closures of either the Redfield- or NIBA-type memory kernels, we first provide an overview of the underlying types of correlation functions that are employed in the calculation of the auxiliary kernels.  These are summarized in Eqs. (\ref{Eq:q00})--(\ref{Eq:q11p}) of Appendix \ref{App:Closures}.  For convenience, we reproduce these expressions, within the Ehfenfest approximation, below, 
	\begin{align}
	q^{(00)}_{nm}(t)  &= \int d\Gamma\ \rho_B^W(0) \mathrm{Tr}_{\mathrm{sys}}[A_n^\dagger A_m(t)], \label{Eq:q00Eh}\\
	q^{(01)}_{nm}(t)  &= \int d\Gamma\ \rho_B^W(0) V_B^W(t) \mathrm{Tr}_{\mathrm{sys}}[A_n^\dagger A_m(t)], \label{Eq:q01Eh}\\
	q^{(10s)}_{nm}(t) &= \int d\Gamma\ \rho_B^W(0) V_B^W(0) \mathrm{Tr}_{\mathrm{sys}}[A_n^\dagger A_m(t)],\label{Eq:q10cEh}\\
	q^{(10a)}_{nm}(t) &= \int d\Gamma\ \rho_B^W(0) \xi^{W}(0) \mathrm{Tr}_{\mathrm{sys}}[A_n^\dagger A_m(t)],\label{Eq:q10pEh}\\
	q^{(11s)}_{nm}(t) &= \int d\Gamma\ \rho_B^W(0) V_B^W(0) V_B^W(t) \mathrm{Tr}_{\mathrm{sys}}[A_n^\dagger A_m(t)],\label{Eq:q11cEh}\\
	q^{(11a)}_{nm}(t) &= \int d\Gamma\ \rho_B^W(0) \xi^{W}(0) V_B^W(t) \mathrm{Tr}_{\mathrm{sys}}[A_n^\dagger A_m(t)],\label{Eq:q11pEh}
	\end{align}
where $V_B^W = \sum_k c_k x_k$ and $\zeta^W = - \alpha \sum_j c_j P_j \tanh(\beta \omega_j / 2) / \omega_j$. 

Inspection of Eqs.~(\ref{Eq:q00Eh})--(\ref{Eq:q11pEh}) reveals that there are two main types of functions containing bath operators: those that require their sampling exclusively at $t = 0$ [Eqs. (\ref{Eq:q00Eh}), (\ref{Eq:q10cEh}) and (\ref{Eq:q10pEh})], and those containing both statically and dynamically sampled bath operators [Eq. (\ref{Eq:q01Eh}), (\ref{Eq:q11cEh}), and (\ref{Eq:q11pEh})].  We recall the important fact that at $t=0$, the Ehrenfest method is exact and the accuracy of the method diminishes with increasing simulation time.\cite{Golosov2001b} What is not clear, however, is whether the accuracy associated with the \textit{dynamical} sampling of a single system operator is the same as that of a product of system and bath operators, as is required in Eqs. (\ref{Eq:q01Eh}), (\ref{Eq:q11cEh}), and (\ref{Eq:q11pEh}).  Because the bath is treated classically, dynamical sampling of bath operators may indeed accrue larger errors. Since the classical approximation is most problematic for high frequency modes, this problem may be exacerbated by fast baths characterized by broad spectral densities, namely the Debye spectral density. Instead, when bath operators are sampled statically, as is the case in Eqs. (\ref{Eq:q00Eh}), (\ref{Eq:q10cEh}), and (\ref{Eq:q10pEh}), the $t=0$ weighting of trajectories of the correlation functions is captured exactly.  One may also distinguish the correlation functions above on the basis of sampling of \textit{distinct} bath operators not normally included, whether explicitly or implicitly, in $\mathcal{C}(t)$. Naively, one may suppose that Eqs.~(\ref{Eq:q01Eh})--(\ref{Eq:q11pEh}) contain information distinct from that in contained in $\mathcal{C}(t)$, but the Ehrenfest evolution algorithm requires sampling of $V_B^W(t)$, which contributes a dynamic component to the system's bias energy $\varepsilon \mapsto \varepsilon + V_B^W(t)$.  Consequently, only Eqs.~(\ref{Eq:q10pEh}) and (\ref{Eq:q11pEh}), which sample $\zeta^W$, contain \textit{distinct} information about the system which is not already included in the calculation of $\mathcal{C}(t)$. Indeed, these correlation functions may contain additional information about the system-bath interaction via statically sampled bath operator, $\zeta^W$, that facilitate the improvement over the bare Ehrenfest dynamics afforded by the memory function approach.


\begin{figure}[t]
\includegraphics[width=8.5cm]{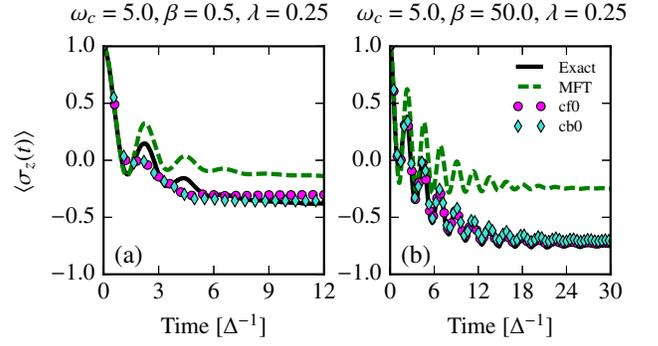} 
\caption{Comparison of the dynamics resulting from the $\mathcal{Q}$-forward and $\mathcal{Q}$-backward $cf0$ (fuschia dots) and $cb0$ (blue diamonds) closures for the Redfield-type memory kernel. Exact results are obtained from Ref. \onlinecite{Thoss2001}.  }\label{Fig:WTMQforQback}
\end{figure}


A distinct source of error that may affect the accuracy of closures that implement numerical time-derivatives lies in the accuracy of the time-derivatives themselves.  If the correlation functions calculated via the Ehrenfest procedure are sufficiently smooth and well converged and the time-step is sufficiently small, this error can be expected to be minimal.  However, correlation functions containing bath operators that require sampling at finite times tend to be highly oscillatory, especially for fast baths, which may lead to less accurate results.  

Armed with these considerations, it is possible to explore the differences associated with the different closures of the auxiliary kernels.   Because the Redfield- and the NIBA-type projectors require different combinations of the aforementioned correlators, Eqs.~(\ref{Eq:q00Eh})--(\ref{Eq:q11pEh}), as input for their auxiliary kernels, we discuss the behaviors of the different closures for the two projectors separately.  

\begin{figure}[b]
\includegraphics[width=8.5cm]{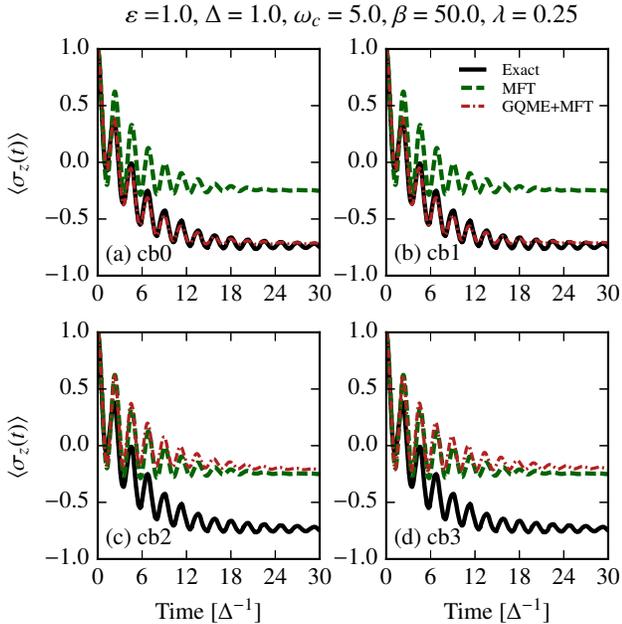} 
\caption{Comparison of population dynamics obtained from the $cb0$, $cb1$, $cb2$, and $cb3$ closures for the Redfield-type kernels, with $\tau_c$ = 2.0. Exact results are obtained from Ref. \onlinecite{Thoss2001}.}\label{Fig:WTMConvPred}
\end{figure}

\begin{figure}[t]
\includegraphics[width=8.5cm]{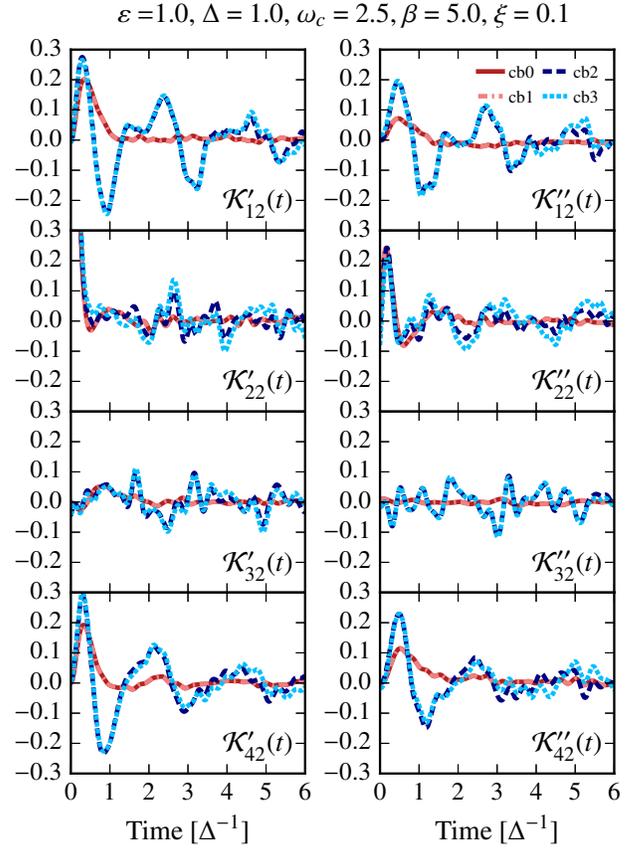} 
\caption{Comparison of Redfield-type memory kernel elements for $\mathcal{Q}$-backward closures $cb0$, $cb1$, $cb2$, and $cb3$ for the SB model with Ohmic spectral density. Consistent with the notation in Figs. \ref{Fig:Red_KMFig3b_Kall} and \ref{Fig:Debye_K}, the memory kernel elements are separated into real and imaginary components, $\mathcal{K}_{x2}(t) = \mathcal{K}'_{x2}(t) + i\mathcal{K}''_{x2}(t)$.}\label{Fig:KernelClosureComparisons}
\end{figure}

\begin{figure}[t]
\includegraphics[width=8.5cm]{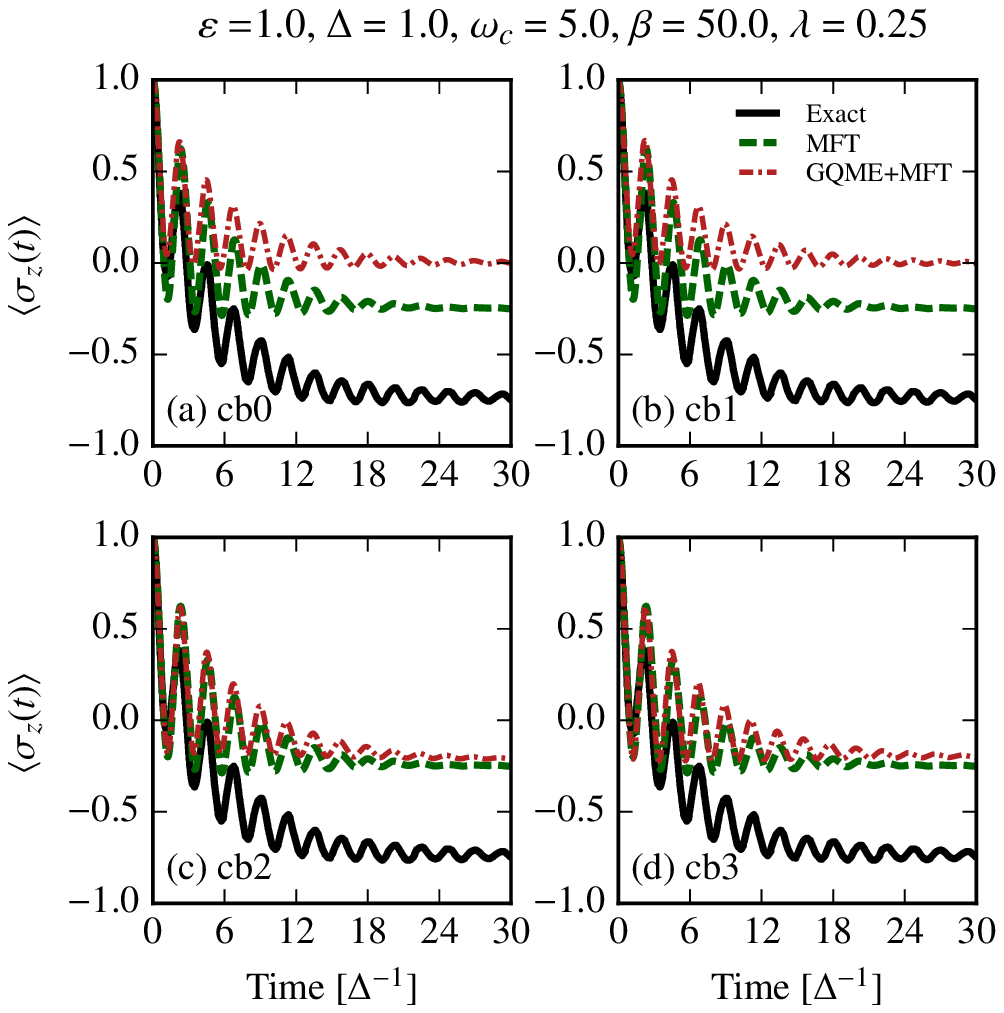} 
\caption{Comparison of population dynamics obtained from the $cb0$, $cb1$, $cb2$, and $cb3$ closures for the Redfield-type kernels when the making the approximation $[\rho_B V_B]^W \approx V_B^W V_B^W$. Exact results are obtained from Ref. \onlinecite{Thoss2001}. }\label{Fig:WTMConvPred_xiOFF}
\end{figure}

Focusing first on the Redfield-type kernel closures, we assess the effect of the 
 $\mathcal{Q}$-backward and $\mathcal{Q}$-forward approaches by focusing first on the
 $cb0$ and $cf0$ closures.  Inspection of Eqs. (\ref{Eq:K3b1}) and (\ref{Eq:K3f1}) reveals that the
 only difference between the $\mathcal{Q}$-forward and $\mathcal{Q}$-backward 
 closures lies in the fact that $\mathcal{K}^{(3f)}(t)$ contains the time-evolved 
 bath operator $V_B^W(t)$, whereas $\mathcal{K}^{(3b)}(t)$ requires sampling of the static bath 
 operator, $V_B^W(0)$. Consistent with the above discussion, we may expect that the 
 $cf0$ closure will lead to less accurate results than $cb0$.  As 
 Fig. \ref{Fig:WTMQforQback} shows, the difference between the two closures is minimal. To understand the smallness of the difference between these two closures, 
 it is sufficient to consider that, while each closure has a different form for $\mathcal{K}^{(3)}(t)$, both closures share the same form for $\mathcal{K}^{(1)}(t)$, which requires the sampling of bath operators both at $t=0$ and at finite times. Hence, any error associated with the 
 explicit inclusion of time evolved bath operators would affect both closures, $cb0$ 
 and $cf0$, and any benefit derived from the exclusive sampling of $t = 0$ bath operators is also maintained in the form for $\mathcal{K}^{(1)}(t)$. Further, the similarity in performance of the $cb0$ and $cf0$ closures indicates that the error associated with the dynamical sampling of products of system and bath operators is often similar to the error associated with the exclusive sampling of system operators. Because the difference between the results of the 
 $\mathcal{Q}$-forward and $\mathcal{Q}$-backward closures is small, we henceforth exclusively address the differences among the $\mathcal{Q}$-backward closures, $cb0$, $cb1$, $cb2$, 
 and $cb3$.

Consideration of the remaining three $\mathcal{Q}$-backward closures, $cb1$, $cb2$, and $cb3$, likewise requires close scrutiny of the types of correlation functions that are used in each.  First we note that, while the form of $\mathcal{K}^{(3b)}(t)$ in the $cb0$ closure avoids the sampling of bath operators at finite time, $\mathcal{K}^{(1)}(t)$ still samples bath operators at finite times.  Instead, the $cb1$ closure completely avoids the sampling of bath operators at $t \neq 0$, while still benefiting from sampling of static bath operators for both auxiliary kernels.  In contrast, the $cb2$ and $cb3$ closures explicitly avoid sampling of static bath operators, other than the density operator for the bath. As is evident from panels (a) and (b) in Fig. \ref{Fig:WTMConvPred}, the $cb1$ closure performs as well or better than the $cb0$ closure. However, because \textit{the $cb1$ closure leads to equations that are easier to converge and auxiliary kernels that are easier to calculate}, it should be preferred over the $cb0$ closure.  

Inspection of panels (c) and (d) of the same figure shows that the $cb2$ and $cb3$ essentially recover the bare Ehrenfest behavior.  Thus, we have demonstrated the remarkable fact that different closures for the memory function, all of which are exact when implemented with exact input, can yield markedly different results when combined with approximate dynamical input. The analytical proof that, for example, the $cb3$ (and $cf3$) closure must yield correlators that are identical to the use of the bare input dynamics is provided in the companion paper.\cite{KellyMontoya2015} 
Further, the fact that the $cb2$ closure also recovers the bare Ehrenfest dynamics clearly indicates that the first of the two criteria specified in Ref.~\onlinecite{KellyMontoya2015} is satisfied by the Ehrenfest method.  Indeed, the numerical data show that within the Ehrenfest approach the action of the Liouvillian acting on a dynamically sampled operator is equivalent to the numerical time-derivative of the analogous correlation function. In addition, given the violation of the second criterion ($[\mathcal{L}^{Eh}, (e^{i\mathcal{L}t})^{Eh}] \neq 0$), it is not surprising that the $cb0$ and $cb1$ closures yield GQME dynamics that are \textit{distinct} from the the results of direct application of the Ehrenfest method.  However, the violation of the second criterion does not explain the reason for marked improvement in the dynamics afforded by the $cb0$ and $cb1$ closures. 
These results also lend additional credence to the claim that the success of the GQME+MFT approach does \textit{not} rely on the short-time accuracy of Ehrenfest dynamics.  Further, they illustrate that improvement within the memory formalism over the bare quasi-classical theory depends sensitively on the correlation functions calculated as input for the auxiliary memory kernels.   In Sec. \ref{Subsec:ProjectionOperators} we suggested that the Ehrenfest method might capture the dynamics of coherences more accurately than that of the populations, and that the self-consistent extraction of the memory kernel would include corrections to the population dynamics afforded by the ostensibly more accurate coherence dynamics. However, the recovery of Ehrenfest dynamics by closures $cb2$ and $cb3$, which also use the dynamics of coherences, implies that this cannot be the root cause of the improvement of dynamics within the memory function approach.  Instead, the more likely explanation is that the sampling of static bath operators in Eqs.~(\ref{Eq:q10cEh}) and (\ref{Eq:q11cEh}) contributes important information about the system-bath interaction, which leads to far greater accuracy in the extracted memory kernels themselves.  


\begin{figure}[b]
\includegraphics[width=8.5cm]{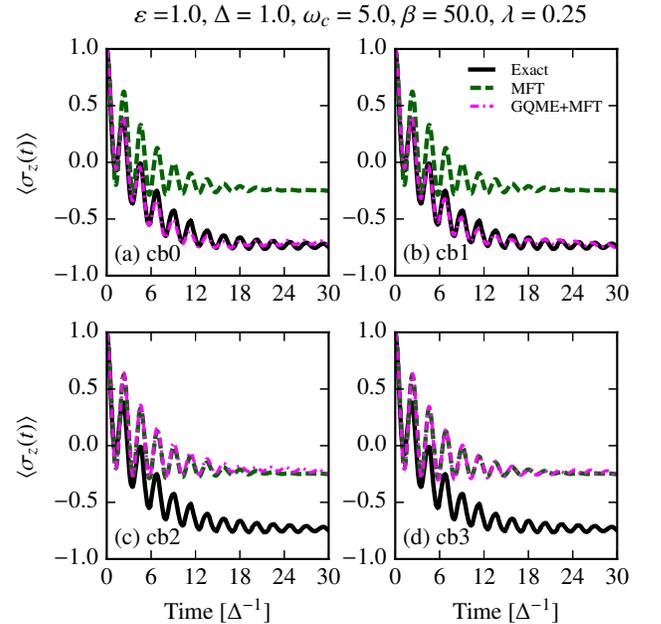} 
\caption{Comparison of population dynamics obtained from the $\mathcal{Q}$-backward closures $cb0$, $cb1$, $cb2$, and $cb3$ for the NIBA-type kernels with $\tau_c = 20.0$. Exact results are obtained from Ref. \onlinecite{Thoss2001}.}\label{Fig:WTMConvPniba}
\end{figure}

The differences in the dynamics resulting from the different closures are also evident in the extracted memory kernels.  Direct comparison of the Redfield-type memory kernels for cases characterized by the Debye spectral density fails to reveal much, since their highly oscillatory behavior obfuscates subtle differences among the memory kernels obtained from different closures.  However, the fast decay of the Ohmic spectral density, which results in quickly decaying memory kernels with minor oscillations, makes discerning qualitative and quantitative differences between the extracted kernels possible.  Fig. \ref{Fig:KernelClosureComparisons} compares a representative set of Redfield-type kernel elements arising from different closures.  As is clear from the figure, all closures agree within numerical and sampling error in their $t = 0$ values.  However, the $cb2$ and $cb3$ closures display a stronger oscillatory behavior, in contrast to the $cb0$ and $cb1$ closures, which correctly recover accurate dynamics (see Fig. \ref{Fig:KMFig3b_BestPopulations}).  We further note that the difference in behavior is greatest at intermediate times.

The previous discussion suggests that the main factor leading to highly accurate memory kernels is the \textit{exact} sampling of specific static bath operators.  A corollary question arises: do all correlation functions with statically sampled bath operators lead to a similar improvement?  After all, Eq.~(\ref{Eq:q01Eh}) samples $V_B^W(0)$ at $t = 0$, but its use in closure $cb3$ does not lead to any improvement over the bare Ehrenfest dynamics. This suggests either that the exact sampling of exclusively static bath operators adds an important correction to the auxiliary kernels, or that correlation functions that sample $V_B^W(0)$ are not as important as those that sample $\zeta^W$. The idea that $\zeta^W$ is neither sampled explicitly nor implicitly (via the Ehrenfest evolution protocol) in the calculation of $\mathcal{C}(t)$ provides some support to the latter claim. It is also fair to ask whether one can similarly benefit from ``improperly'' Wigner-transformed bath operator products sampled at $t=0$. This question becomes particularly important when a functional form for the density operator of the bath is either not available or challenging to obtain. To see the importance of properly including the terms in the Wigner transformation, we take an approximate form for the Wigner transform of the product $\rho_B V_B$ (defined in Appendix \ref{App:EhrenfestCFs}). Our approximation truncates the Moyal expansion for the Wigner transform of a product of operators at zeroth order in $\hbar$, neglecting the second term (containing $\zeta^W$) on the right side of Eq.~(\ref{Eq:WignerTransformrhoVB}).  Results for this approximation are shown in Fig.~\ref{Fig:WTMConvPred_xiOFF}.  As is evident in the figure, the benefits in closures $cb0$ and $cb1$ that originally led to the quantitative agreement between the GQME+MFT and exact dynamics are eliminated.  Instead, the final result, while not unphysical, is not better than the standard Ehrenfest result.  Since the $cb2$ and $cb3$ closures do not contain this neglected term, the results in panels (c) and (d) of Fig. \ref{Fig:WTMConvPred_xiOFF} are the same as those in panels (c) and (d) of Fig. \ref{Fig:WTMConvPred}. This result underscores the importance of proper sampling of \textit{all} contributions arising from the Wigner transform of operator products.  

\begin{figure}[t]
\includegraphics[width=8.5cm]{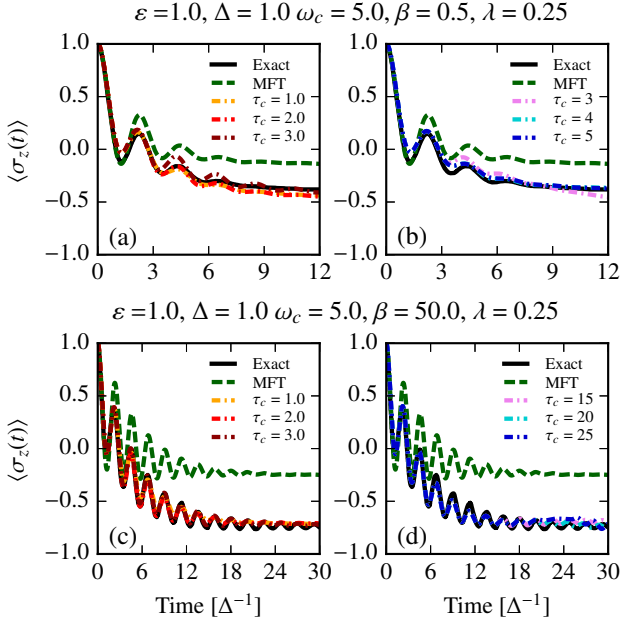} 
\caption{Comparison of population dynamics obtained using the $cb1$ closures for the Redfield- and NIBA-type kernels with varying $\tau_c$. Panels (a) and (c) correspond to the Redfield-type projector, while panels (b) and (d) correspond to the NIBA-type projector. Exact results are obtained from Ref. \onlinecite{Thoss2001}.}\label{Fig:CompTau}
\end{figure}

\begin{figure}[t]
\includegraphics[width=8.5cm]{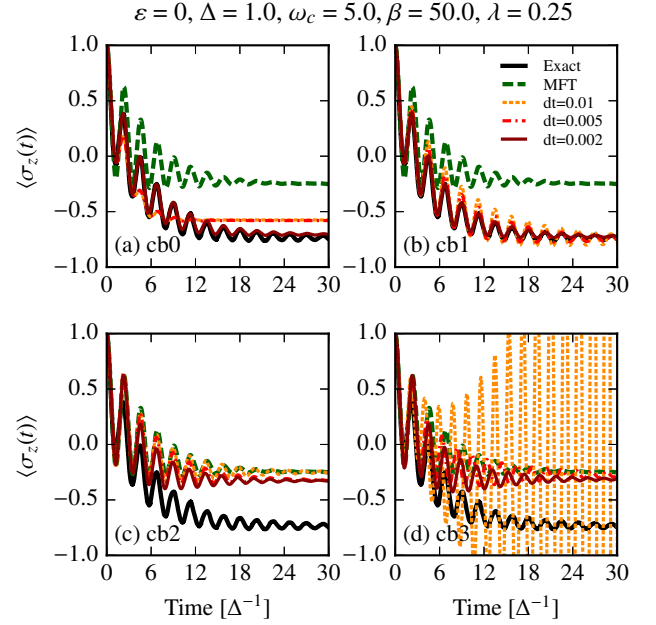} 
\caption{Comparison of population dynamics obtained from the $\mathcal{Q}$-backward closures $cb0$, $cb1$, $cb2$, and $cb3$ for the Redfield-type kernels with $\tau_c = 2.0$ upon varying the time step used in the GQME evolution. Note that the actual time step used in the GQME solution is twice that used in the calculation of the memory kernels. Exact results are obtained from Ref. \onlinecite{Thoss2001}.}\label{Fig:WTMConvPred_dtGQME}
\end{figure}

In comparison to the Redfield-type closures, the NIBA-type kernels only contain two types of correlation functions, $q^{(00)}(t)$ and $q^{(10a)}(t)$. As mentioned in the discussion of the Redfield-type closures, $q^{(00)}(t)$ contains the same information as Ehrenfest version of $\mathcal{C}(t)$, whereas $q^{(10a)}(t)$ contains \textit{exact} information about the system-bath interaction at $t = 0$.  For this reason, we expect the $c0b$ and $c1b$ closures to yield significantly better dynamics than the $cb2$ and  $cb3$ closures, which should simply recover the Ehrenfest dynamics.\cite{KellyMontoya2015}  Indeed, as Fig.~\ref{Fig:WTMConvPniba} shows, the $cb0$ and $cb1$ closures are able to quantitatively correct the Ehrenfest dynamics. In contrast to the Redfield case, the $cb0$ and $cb1$ NIBA-type auxiliary kernels require the sampling of the same bath operators, which explains the lack of difference in the behaviors of the two closures.  

Of paramount importance to the success of the memory approach is the finite lifetime of the memory kernel.  For the GQME+MFT implementation used here, we have chosen a cutoff time for the memory kernel, $\tau_c$, which lies in a stability plateau alluded to in Sec. \ref{Sec:Results}.  The range of the stability plateau can depend sensitively on the regime of parameter space explored.  Fig. \ref{Fig:CompTau} shows the dependence of the GQME+MFT dynamics for the Redfield- and NIBA-type projectors on the specific value of $\tau_c$ used.  Panels (a) and (b), corresponding to a fast bath, high temperature, biased case, show the greatest sensitivity of the GQME dynamics to the exact cutoff time, $\tau_c$.  In contrast, the results in panels (c) and (d), which correspond to lower temperatures, are more stable.  Despite the slight sensitivity of the results on the choice of $\tau_c$ for the examples shown, the GQME dynamics are clearly robust. We remark, however, that there are regions of parameter space for which this stability plateau is short-lived or nonexistent.  In those cases, the present approach is clearly not appropriate. 

Finally, we address the convergence properties of the GQME dynamics with respect to the time-step of the memory kernel and the GQME evolution algorithm. As Fig. \ref{Fig:WTMConvPred_dtGQME} shows, decreasing the time step used in the extraction of the memory kernel can greatly alter the accuracy of the GQME dynamics.  The closures most sensitive to the time step are $cb0$ and $cb3$ closures.  The sensitivity of the $cb3$ closure to the time step is not difficult to understand, since for large $dt$ the second numerical time derivative becomes noisy, leading to the growing oscillations in panel (d).  Because closure $cb0$ shown in panel (a) contains the correlation functions $q^{(11s)}(t)$ and $q^{(11a)}(t)$, which require sampling of bath operators at $t = 0$ and at finite times and are highly oscillatory functions, a smaller time step is required to achieve sufficiently accurate memory kernels.  It is important to note that, as stated before, the $cb0$ and $cb1$ closures, when fully converged with respect to bath realizations and time step, yield identical results that greatly improve the Ehrenfest results, while closures $cb2$ and $cb3$ are only capable of recovering the Ehrenfest dynamics.

\section{Conclusions}
\label{Sec:Conclusions}

In this paper we have developed a method to obtain the nonequilibrium population and coherence dynamics based on the Mori formalism.  Our approach is general and, depending on the choice of projector, can treat arbitrary single-time nonequilibrium populations and coherences as well as more complicated dynamical objects, such as multi-time, equilibrium and nonequilibrium correlation functions. We have shown that use of the Redfield-type projector recovers the conventional NZ treatment previously used by Shi and Geva\cite{Shi2003, Shi2004, Zhang2006a} and Kelly, Markland, and coworkers\cite{Kelly2013, Kelly2015, Pfalzgraff2015} in the context of the SB model, and Rabani and coworkers\cite{Cohen2011, Cohen2013, Wilner2013, Wilner2014, Wilner2014a} for more general models. 


While previous applications of the GQME+semi-classical approach\cite{Shi2004, Kelly2013, Kelly2015} have been limited to the Redfield-type projector and have focused on the improvement over the bare semi-classical dynamics that the memory function formalism can afford, we have systematically explored the sensitivity of the results to the choice of projector and the type of closure employed. In doing so, we find two important facts. First, slowly decaying memory kernels, often observed when using the NIBA-type projector, do not result in an inaccurate description of the GQME dynamics. This demonstrates that the success of the GQME+semi-classical approach is \textit{not} a function of the short-time accuracy of the approximate method used to calculate the auxiliary kernels. Second, we identify the types of closures that consistently lead to improvements over the bare semi-classical dynamics ($cb0$, $cb1$, $cf0$, and $cf1$) and attribute this improvement, in part, to the sampling of static bath operators, $\zeta^W$, which do not appear in the evaluation of the approximate bare populations. Just as importantly, we identify the types of closures that recover the Ehrenfest dynamics ($cb2$, $cb3$, $cf2$, and $cf3$).  Our findings also provide numerical confirmation of the analytical proof included in the companion paper,\cite{KellyMontoya2015} which indicates that use of the $cb3$ and $cf3$ closures \textit{can only} recover the level of dynamics used to calculate the auxiliary kernels.  

Finally, we remark that the Mori-based formulation furthered in this work provides a flexible framework to accurately study problems that go beyond the scope of nonequilibrium dynamics for SB-type models. For instance, the Mori formalism can easily address equilibrium and multi-time correlation functions in systems coupled to harmonic \textit{and} anharmonic baths, as well as problems where the system-bath distinction is absent. Work in this latter direction will be pursued in future papers.

\section{Acknowledgements}

The authors would like to thank Aaron Kelly and Tom Markland for helpful discussions.  D.R.R. acknowledges support from NSF No. CHE-1464802. A.M.C. thanks Hsing-Ta Chen and Guy Cohen for useful conversations.

\appendix

\section{Fourier-Laplace Analysis of Closures}
\label{App:LFTAnalysis}

We begin by introducing the Fourier-Laplace transform of a time-dependent function $f(t)$,  
	\begin{equation}
	f(\omega) = \int_0^{\infty} dt e^{i\omega t} f(t). 
	\end{equation}
Its first and second time-derivatives take the following form, 
	\begin{align}
	\int_0^{\infty} dt e^{i\omega t} \dot{f}(t) &= -f(0) + (-i\omega ) f(\omega),\\
	\int_0^{\infty} dt e^{i\omega t} \ddot{f}(t) &= -\dot{f}(0) - (-i\omega)f(0) + (-i\omega)^2 f(\omega).
	\end{align}
	
The kernel expansions in Eqs. (\ref{Eq:SCKb}) and (\ref{Eq:SCKf}) can be rewritten as follows, 
	\begin{align}
	\mathcal{K}(\omega) &= [1 - \mathcal{K}^{(3b)}(\omega)]^{-1}\mathcal{K}^{(1)}(\omega) \label{Eq:Kwb0}\\
	&= \mathcal{K}^{(1)}(\omega)[1 - \mathcal{K}^{(3f)}(\omega)]^{-1}.\label{Eq:Kwf0}
	\end{align}


In their original paper, Shi and Geva\cite{Shi2003} derived the following identity for the Redfield-type (thermal) projector, 
	\begin{align}
	\mathcal{Q}\mathcal{L}\mathcal{Q} &= \mathcal{Q}(\mathcal{L} - \mathcal{P}\mathcal{L}_{sb}),\\
	&= (\mathcal{L} - \mathcal{L}_{sb}\mathcal{P})\mathcal{Q},
	\end{align}
where $\mathcal{L}_{sb}$ is the Liouvillian corresponding to the system-bath interaction for the SB model $V = \alpha \sigma_z\sum_k c_k x_k$.   The second line is a simple extension of the derivation provided by Shi and Geva.  These identities allow for the exact rewriting of the ``$\mathcal{Q}$-surrounded'' projected propagator, 
	\begin{align}
	\mathcal{Q} e^{i\mathcal{Q}\mathcal{L}t}\mathcal{Q} &=  \mathcal{Q} e^{i(\mathcal{L} - \mathcal{P}\mathcal{L}_{sb})t}\mathcal{Q}\\
	&=  \mathcal{Q} e^{i(\mathcal{L} - \mathcal{L}_{sb}\mathcal{P})t}\mathcal{Q}.
	\end{align}
Replacing these expressions for the projected propagator in Eq. (\ref{Eq:MemoryKernel2}) followed by use of the Dyson decomposition leads to the following \textit{three}- rather than \textit{two}-membered expansions,
	\begin{align}
	\mathcal{K}(t) &= \mathcal{K}^{(1)}(t) + \int_0^t d\tau\  \mathcal{K}^{(2b)}(t-\tau)\mathcal{K}^{(1)}(\tau), \\
	&= \mathcal{K}^{(1)}(t) + \int_0^t d\tau\  \mathcal{K}^{(1)}(t-\tau)\mathcal{K}^{(2f)}(\tau) ,
	\end{align}
where the second auxiliary kernel also contains the projected propagator, 
	\begin{align}
	\mathcal{K}_{nm}^{(2b)}(t) &= -i \langle \bra{\rho_B A_n} \mathcal{L}_{sb} e^{i(\mathcal{L} - \mathcal{P}\mathcal{L}_{sb})t} \ket{A_m} \rangle, \\
	\mathcal{K}_{nm}^{(2f)}(t) &= -i \langle \bra{\rho_B A_n} e^{i(\mathcal{L} - \mathcal{L}_{sb}\mathcal{P})t} \mathcal{L}_{sb} \ket{A_m} \rangle. 
	\end{align}
In Fourier-Laplace space, the previous expressions for the memory kernel become, 
	\begin{align}
	\mathcal{K}(\omega) &= \mathcal{K}^{(1)}(\omega) +  \mathcal{K}^{(2b)}(\omega)\mathcal{K}^{(1)}(\omega), \label{Eq:Kwb}\\
	&= \mathcal{K}^{(1)}(\omega) +  \mathcal{K}^{(1)}(\omega)\mathcal{K}^{(2f)}(\omega) .\label{Eq:Kwf}
	\end{align}

In the time-domain, the second auxiliary kernels may be expanded as follows,
	\begin{align}
	\mathcal{K}^{(2x)}(t) &= \mathcal{K}^{(3x)}(t) + \int_0^t d\tau\  \mathcal{K}^{(2x)}(t-\tau)\mathcal{K}^{(3x)}(\tau), \label{Eq:K2x1}\\
	&= \mathcal{K}^{(3x)}(t) + \int_0^t d\tau\  \mathcal{K}^{(3x)}(t-\tau)\mathcal{K}^{(2x)}(\tau) ,\label{Eq:K2x2}
	\end{align}
where $x \in \{ b, f\}$.  Transforming Eqs. (\ref{Eq:K2x1}) and (\ref{Eq:K2x2}), we may solve for $\mathcal{K}^{(2x)}(\omega)$
	\begin{align}
	\mathcal{K}^{(2x)}(\omega) &= \mathcal{K}^{(3x)}(\omega)[1 +  \mathcal{K}^{(3x)}(\omega)]^{-1}, \label{Eq:K2xw1}\\
	&= [1 +  \mathcal{K}^{(3x)}(\omega)]^{-1} \mathcal{K}^{(3x)}(\omega).\label{Eq:K2xw2}
	\end{align}
Substitution of Eqs. (\ref{Eq:K2xw1}) and (\ref{Eq:K2xw2}) into Eqs. (\ref{Eq:Kwb}) and (\ref{Eq:Kwf}) for the $\mathcal{Q}$-forward and $\mathcal{Q}$-backward closures yields, 
	\begin{align}
	\mathcal{K}(\omega) &= [1 - \mathcal{K}^{(3b)}(\omega)]^{-1}\mathcal{K}^{(1)}(\omega)\\
	&= \mathcal{K}^{(1)}(\omega)[1 - \mathcal{K}^{(3f)}(\omega)]^{-1},
	\end{align}
which are clearly equivalent to Eqs. (\ref{Eq:Kwb0}) and (\ref{Eq:Kwf0}), implying that the three-membered closures are equivalent to the two-membered closures.

\section{Expressions for Auxiliary Memory Kernels}
\label{App:Closures}

Here we provide explicit expressions for the components of the memory kernel using the Redfield- and NIBA-type projection operators.  Before going further, however, we introduce for notational clarity the following correlation functions.
	\begin{align}
	q^{(00)}_{nm}  &= \mathrm{Tr}[\rho_B A_n^{\dagger} A_m(t)], \label{Eq:q00}\\
	q^{(01)}_{nm}  &= \mathrm{Tr}[\rho_B A_n^{\dagger} A_m(t)V_B(t)],\label{Eq:q01}\\
	q^{(10s)}_{nm} &= \frac{1}{2}\mathrm{Tr}[\{V_B,\rho_B\} A_n^{\dagger} A_m(t)],\label{Eq:q10c}\\
	q^{(10a)}_{nm} &=\frac{-i}{2} \mathrm{Tr}[[V_B,\rho_B] A_n^{\dagger} A_m(t)],\label{Eq:q10p}\\
	q^{(11s)}_{nm} &= \frac{1}{2}\mathrm{Tr}[\{V_B,\rho_B\} A_n^{\dagger} A_m(t)V_B(t)],\label{Eq:q11c}\\
	q^{(11a)}_{nm} &=\frac{-i}{2} \mathrm{Tr}[[V_B,\rho_B] A_n^{\dagger} A_m(t)V_B(t)],\label{Eq:q11p}
	\end{align}
	where $s$ and $a$ indicate symmetrized (anticommutator) or antisymmetrized (commutator) bath products.

Using the Redfield-type projector, 
	\begin{equation}\label{Eq:RedfieldProjector}
	\mathcal{P}_{Red} = \sum_i\ket{A_i}\rangle \langle\bra{\rho_B A_i}, 
	\end{equation}
where $A_i \in \{\ket{0}\bra{0}, \ket{1}\bra{0}, \ket{1}\bra{0}, \ket{1}\bra{1} \}$, $i \in \{1,2,3,4\}$, and $\rho_B = e^{-\beta H_B}/ \mathrm{Tr}_B[e^{-\beta H_B}]$, the elements of the memory kernels take the following forms, 
	\begin{align}\label{Eq:RP}
	[\mathcal{K}^{(1)}(t)]_{nm} &= [X_{n} q_{nm}^{(11s)}(t) + i Y_{n}q_{nm}^{(11a)}(t)] X_{m}, \\ 
	[\mathcal{K}^{(3b)}(t)]_{nm} &= iX_{n} q_{nm}^{(10s)}(t) -  Y_{n}q_{nm}^{(10a)}(t), \\
	[\mathcal{K}^{(3f)}(t)]_{nm} &= iq_{nm}^{(01)}(t) X_{m},
	\end{align}
where $X_n = 2(\delta_{n3} - \delta_{n2})$ and $Y_n = 2 (\delta_{n1} - \delta_{n4})$.

Using the NIBA-type projector, 
	\begin{equation}\label{Eq:mixNIBAProjector}
	\mathcal{P}_{NIBA} = \sum_i\ket{B_i}\rangle \langle\bra{\rho_B B_i}, 
	\end{equation}
where $B_i \in \{\ket{0}\bra{0}, \ket{1}\bra{1} \}$, $i \in \{1,2\}$ and $\rho_B = e^{-\beta H_B}/ \mathrm{Tr}_B[e^{-\beta H_B}]$, the elements of the memory kernels take the following forms, 
	\begin{align}\label{Eq:RP}
	[\mathcal{K}^{(1)}(t)]_{nm} &= 2(-1)^{n+m}\Big[2\Delta \Im[q_{n^22}^{(10a)}(t)] \nonumber \\
	& \qquad - \Delta^2 \Re[  q_{32}^{(00)}(t) - q_{33}^{(00)}(t)]\Big],\\
	[\mathcal{K}^{(3b)}(t)]_{nm} &= 2  (-1)^n \Big[ q_{n^2m^2}^{(10a)}(t) +\Delta \Im[ q_{n^22}^{(00)}(t)] \Big], \\
	[\mathcal{K}^{(3f)}(t)]_{nm} &= 2 \Delta (-1)^m \Im[ q_{n^22}^{(00)}(t)].
	\end{align}
We employ a notation where some indices are squared since the $q(t)$ functions are labelled using the indices corresponding to the $A_j$ operators, which are related to the $B_j$ operators in the following way: $B_1 \mapsto A_1 = A_{1^2}$ and $B_2 \mapsto A_4 = A_{2^2}$. 

For the projector above to be truly of NIBA-type, $\mathcal{K}^{(1)}(t)$ should be $\mathcal{O}(\Delta^2)$. Instead, $\mathcal{K}^{(1)}(t)$ has contributions of first and second order in $\Delta$.  Indeed, the proper NIBA-type projector has the following form, 
\begin{equation}\label{Eq:NIBAProjector}
	\mathcal{P}_{NIBA} = \sum_i\ket{B_i}\rangle \langle\bra{\rho_B^{(i)} B_i},
	\end{equation}
where $\rho_B^{(i)} = e^{-\beta(H_B - (-1)^i \alpha V_B)} / \mathrm{Tr}_B[e^{-\beta(H_B - (-1)^i \alpha V_B)}]$. 
\section{Initial Conditions in the Ehrenfest method}
\label{App:ICEhrenfest}

In an open quantum system where a subsystem interacts weakly with a heat bath, the Ehrenfest method\cite{Gerber1982, Stock1995,Tully1998a,Grunwald2009} treats the subsystem quantum mechanically and the bath classically.  The validity of this approximation relies on two important assumptions: correlations between the system and bath are negligible, and the characteristic energy of the bath is smaller than the other energy scales in the problem, justifying the use of classical mechanics for the evolution of the bath. 

The Ehrenfest method has been derived from complementary wavefunction\cite{Tully1998a} and density matrix formulations.\cite{Grunwald2009}  Because of its clarity, the derivation based on the density matrix and the quantum-classical Liouville equation has garnered much attention in the last decade. The density matrix formulation only requires that the subsystem and bath density matrices have norms equal to unity.  However, the lack of restriction of the subsystem density matrix to pure states results in ambiguities in its implementation. To illustrate the source of the ambiguity, we focus on the following correlation function for the spin-boson model 
	\begin{equation} \label{Eq:ProblemCF}
	C(t) = \mathrm{Tr}[R_B \sigma_i \sigma_z(t)],
	\end{equation}
where $i \in \{ x, y \}$, and the bath initial condition has unit trace, $\mathrm{Tr}_B[R_B] = 1$.  This corresponds to a nonequilibrium initial condition where the system is initially in a superposition of coherences.  Immediately it is clear that $\rho_S(0) = \sigma_i$ has zero norm.  To remedy this, one may take advantage of the linearity of the problem and rewrite Eq.~(\ref{Eq:ProblemCF}) into sums of correlation functions with proper initial conditions, 
	\begin{align}
	C(t) &= \mathrm{Tr}[R_B (\ket{1}\bra{1} + \sigma_i) \mathcal{O}(t)] - \mathrm{Tr}[R_B \ket{1}\bra{1} \mathcal{O}(t)],  \label{Eq:probCF1} \\
	&= \mathrm{Tr}[R_B (\ket{2}\bra{2} + \sigma_i) \mathcal{O}(t)] - \mathrm{Tr}[R_B \ket{2}\bra{2} \mathcal{O}(t)], \label{Eq:probCF2}\\
	&= \frac{1}{2}\mathrm{Tr}[R_B (\mathbf{1}_S + 2\sigma_i) \mathcal{O}(t)] - \frac{1}{2}\mathrm{Tr}[R_B \mathbf{1}_S \mathcal{O}(t)] \label{Eq:probCF3},
	\end{align}
where $\mathbf{1}_S = \ket{1}\bra{1} + \ket{2}\bra{2}$.

\begin{figure}[t]
\includegraphics[width=8.5cm]{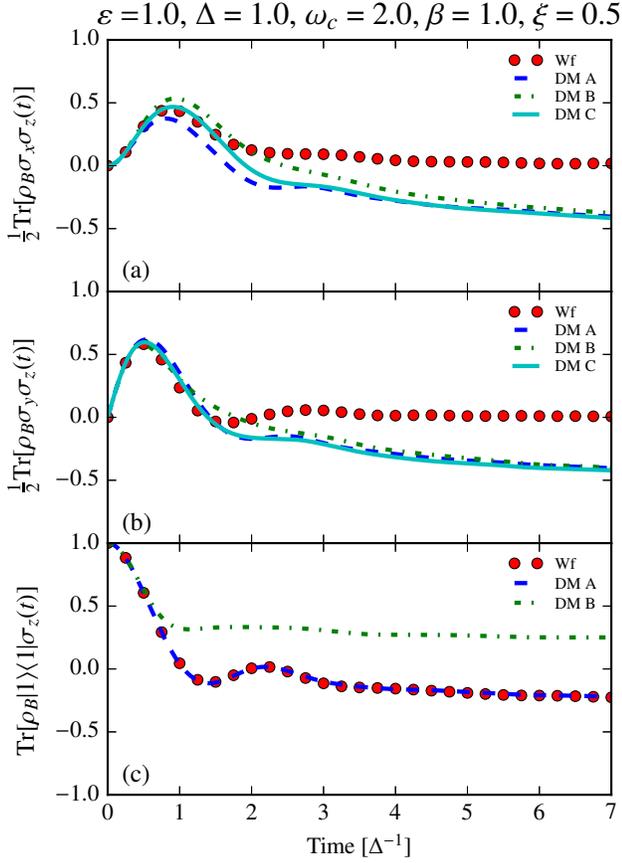} 
\caption{Comparison of Ehrenfest dynamics resulting from the wavefunction- and density matrix-based approaches.  
In panels (a) and (b), DM 1 corresponds to the following combination of initial conditions $[\ket{1}\bra{1} + \sigma_i] - [\ket{1}\bra{1}]$, DM 2 to $[\ket{2}\bra{2} + \sigma_i] - [\ket{2}\bra{2}]$, and DM 3 to $[0.5 * \mathbf{1}_S + \sigma_i] - [0.5 * \mathbf{1}_S]$, where $i = x,y$ for panels (a) and (b), respectively. In panel (c), DM 1 corresponds to $[\ket{1}\bra{1} + \sigma_y] - [\ket{1}\bra{1}]$, and DM 2 to $2 * [0.5 * \mathbf{1}_S] - [\ket{1}\bra{1}]$. }\label{Fig:ICs}
\end{figure}	

Returning to the wavefunction-based derivation of the Ehrenfest approach requires that any subsystem initial condition correspond to a pure state. Referring again to Eq.~(\ref{Eq:ProblemCF}), we may rewrite the trace over the subsystem in the eigenbasis of the subsystem's initial condition, 
	\begin{equation}\label{Eq:probCF4}
	C(t) = \sum_i \lambda_i \mathrm{Tr}_B[R_B \bra{\lambda_i} \sigma_z(t) \ket{\lambda_i}].
	\end{equation}
Although there are other ways of choosing a pure state so as to evaluate the above correlation function, the wavefunction formulation avoids the ambiguity fostered by the density matrix derivation. 

Fig.~\ref{Fig:ICs} shows the calculation of the nonequilibrium population dynamics given different system initial conditions, $\rho_S(0) \in \{ \sigma_x, \sigma_y, \ket{1}\bra{1} \}$.  For the initial conditions corresponding to the Pauli matrices, we implement three different decompositions given by Eqs.~(\ref{Eq:probCF1}), (\ref{Eq:probCF2}), and (\ref{Eq:probCF3}), labeled $A$, $B$, and $C$, respectively.  We also include the results for two different decompositions for the initial condition $\ket{1}\bra{1}$, labeled $A$ and $B$.  As is clear from panels (a) and (b), all density matrix based decompositions agree in their short- and long-time limits, but disagree in their descriptions of intermediate-time behavior.  More importantly, all density matrix decompositions disagree with the wavefunction-based result.  Panel (c) shows the ability of the density matrix-based approach to recover the wavefunction-based result when the initial condition corresponds to a population rather than a coherence, but decomposition $B$ underscores the problems associated with the lack of uniqueness in the density matrix based approach.  

\section{Ehrenfest Method: Correlation Functions}
\label{App:EhrenfestCFs}

Unlike previous implementations of the Ehrenfest method, we are interested in two time correlation functions rather than nonequilibrium single quantity dynamics, 
	\begin{equation}
	C_{AB}(t) = \mathrm{Tr}[\mathcal{A}_S(0) \mathcal{A}_B(0) \mathcal{B}_S(t) \mathcal{B}_B(t)],   
	\end{equation}
where $X_S$ ($X_B$) is a generic system (bath) operator.  Under the quasi-classical approximation of Wigner dynamics, we may rewrite the above correlation function as follows, 
	\begin{equation}
	C_{AB}(t) \approx \int d\Gamma\ \mathcal{A}_B ^W \mathcal{B}^W_B(t)\mathrm{Tr}_S[ \mathcal{A}_S(0)   \mathcal{B}_S(t)],   
	\end{equation}
where the superscript $W$ denotes the Wigner transform of the operator and $\Gamma$ is the set of all classical variables.  

For the auxiliary memory kernels of the spin-boson model, the following Wigner transforms are necessary, 
	\begin{equation}\label{Eq:WignerDistribution}
	\begin{split}
	\rho_B^W &= \prod_j \frac{\tanh(\beta \omega_j / 2)}{\pi} \\
	&\qquad \qquad \exp\Bigg[ -\frac{\tanh(\beta \omega_j / 2)}{\omega_j}  \Big[P_j^2 + \omega_j^2 Q_j^2 \Big]\Bigg],
	\end{split}
	\end{equation}		
	\begin{equation}\
	V_B^W = \alpha \sum_j c_j Q_j,
	\end{equation}
	and, using the Moyal bracket for products of operators,\cite{Imre1967} 
	\begin{equation}
	\begin{split}\label{Eq:WignerTransformrhoVB}
	[V_B \rho_B]^W &= \alpha V_B^W \rho_B^W + i \zeta^W \rho_B^W, 
	\end{split}
	\end{equation}	
	where 
	\begin{equation}
	\zeta^W = - \alpha \sum_j c_j P_j \frac{\tanh(\beta \omega_j / 2)}{\omega_j}
	\end{equation}

Using the above definitions, the correlation functions in Eqs. (A1)--(A6) take the following form under the Ehrenfest approximation, 
	\begin{align}
	q^{(00)}_{nm}(t)  &= \int d\Gamma\ \rho_B^W \mathrm{Tr}_{\mathrm{sys}}[A_n^\dagger A_m(t)],\\
	q^{(01)}_{nm}(t)  &= \int d\Gamma\ \rho_B^W V_B^W(t) \mathrm{Tr}_{\mathrm{sys}}[A_n^\dagger A_m(t)],\\
	q^{(10s)}_{nm}(t) &= \int d\Gamma\ \rho_B^W V_B^W(0) \mathrm{Tr}_{\mathrm{sys}}[A_n^\dagger A_m(t)],\\
	q^{(10a)}_{nm}(t) &= \int d\Gamma\ \rho_B^W \xi^{W}(0) \mathrm{Tr}_{\mathrm{sys}}[A_n^\dagger A_m(t)],\\
	q^{(11s)}_{nm}(t) &= \int d\Gamma\ \rho_B^W V_B^W(0) V_B^W(t) \mathrm{Tr}_{\mathrm{sys}}[A_n^\dagger A_m(t)],\\
	q^{(11a)}_{nm}(t) &= \int d\Gamma\ \rho_B^W \xi^{W}(0) V_B^W(t) \mathrm{Tr}_{\mathrm{sys}}[A_n^\dagger A_m(t)].
	\end{align}

The above considerations regarding the subtlety in the density matrix picture with regard to the implementation of the Ehrenfest method underlines an important interpretation issue.  While it is often regarded that in the Ehrenfest method the system (bath) evolves under the mean field of the classical (quantum) variables, it is important to add the caveat that these mean fields correspond to single rather than ensembles of trajectories.  As such, under the Ehrenfest approximation the system evolves under the time-dependent Hamiltonian defined as
	\begin{equation}
	H_{S, Eh}(t) = [\varepsilon + \lambda^{cl}(t)] \sigma_z + \Delta \sigma_x
	\end{equation}
where the classical bath provides a fluctuating contribution to the bias energy $\lambda^{cl}(t) = \alpha \sum_k c_k Q_k(t)$ and the equation of motion for the density matrix of the system is the Liouville equation using the modified Hamiltonian,
	\begin{equation}
	\frac{d}{dt} \rho_S(t) = -i[H_{S}^{Eh}, \rho(t)].
	\end{equation}
The bath, in turn, evolves under the influence of the time-dependent Hamiltonian
	\begin{equation}
	H_{B}^{Eh}(t) = \frac{1}{2} \sum_k \Big[ P_k^2 + \omega_k^2 Q_k + 2\alpha \bar{\sigma}_z(t) c_k Q_k \Big], 
	\end{equation}
where $\bar{\sigma}_z(t) = \mathrm{Tr}_S[\rho_S(t) \sigma_z]$, and the equations of motion for the classical variables are given by Hamilton's equations,
	\begin{align}
	\frac{dP_k}{dt}  &= - \frac{\partial H_{B}^{Eh}}{\partial Q_k},\\
	\frac{dQ_k}{dt}  &=  \frac{\partial H_{B}^{Eh}}{\partial P_k}.\\
	\end{align}	 

To calculate auxiliary memory kernels used in the present work, trajectories corresponding to a set of initial conditions given by the Wigner distribution, Eq.~(\ref{Eq:WignerDistribution}), are calculated via a second-order Runge-Kutta scheme.  During individual time steps, $\bar{\sigma}_z(t)$ is kept constant for the evolution of the bath, while $\lambda^{cl}(t)$ is kept constant during the evolution of the system. Over a half time step, the equations for the classical variables take the forms,
	    \begin{equation}
    \begin{split}
    Q_{k}\left(t + \frac{\delta t}{2} \right) &= \gamma_{k}(t)\cos\left( \frac{\omega_{k} \delta t}{2} \right) - \frac{\alpha c_{k}}{\omega_{k}^2}\bar{\sigma_z}(t)\\
    &\hspace{1em} + \frac{P_{k}(t)}{\omega_{k}}\cos\left(\frac{\omega_{k} \delta t}{2}\right) ,
    \end{split}
    \end{equation}
and 

    \begin{equation}
    \begin{split}
    P_{k}\left(t + \frac{\delta t}{2} \right) &= P_{k}(t)\cos\left(\frac{\omega_{k} \delta t}{2} \right)+ \omega_{k}\gamma_{k}(t)\sin\left(\frac{\omega_{k} \delta t}{2}\right),
    \end{split}
    \end{equation}
where
    \begin{equation}
    \gamma_{k}(t) =  Q_{k}(t) + \frac{\alpha c_{k}}{\omega_{k}^2}\bar{\sigma_z}(t).
    \end{equation} 
While convergence for correlation functions of system operators only requires only $\sim 10^3-10^4$ trajectories, correlation functions with bath operators require $\sim 3 \times 10^4 - 10^5$ trajectories for sufficiently accurate results.

%


\end{document}